# Thermoresponsive stiffening with microgel particles in a semiflexible fibrin network

Gaurav Chaudhary[1,4], Ashesh Ghosh [2,4], Ashwin Bhardwaj [1,4], Jin Gu Kang [3,4,7], Paul V. Braun [1-5], Kenneth S. Schweizer [2-6], Randy H. Ewoldt [1,4,5] *

[1] Department of Mechanical Science and Engineering, University of Illinois at Urbana-Champaign, Urbana, IL, 61801

[2] Department of Chemistry, University of Illinois at Urbana-Champaign, Urbana, IL, 61801

[3] Department of Materials Science and Engineering, University of Illinois at Urbana-Champaign, Urbana, IL, 61801

[4] Materials Research Laboratory, University of Illinois at Urbana-Champaign, Urbana, IL, 61801, USA

[5] Beckman Institute for Advanced Science and Technology, University of Illinois at Urbana-Champaign, Urbana, IL, 61801, USA

[6] Department of Chemical & Biomolecular Engineering, University of Illinois at Urbana-Champaign, Urbana, IL, 61801, USA

[7] Nanophotonics Research Center, Korea Institute of Science and Technology, Seoul 02792, South Korea

*corresponding author: R.H. Ewoldt (ewoldt@illinois.edu)

**Abstract**

We report temperature-responsive soft composites of semiflexible biopolymer networks (fibrin) containing dispersed microgel colloidal particles of poly(N-isopropylacrylamide) (pNIPAM) that undergo a thermodynamically driven de-swelling transition above a Lower Critical Solution Temperature (LCST). Unlike standard polymer-particle composites, decreasing the inclusion volume of the particles (by increasing temperature) is concomitant with a striking *increase* of the overall elastic stiffness of the composite. We observe such a behavior over a wide composition space. The composite elastic shear modulus reversibly stiffens by up to 10-fold over a small change in temperature from 25-35°C. In isolation, the fibrin network and microgel suspension both soften with increased temperature, making the stiffening of the composites particularly significant. We hypothesize that stiffening is caused by contracting microgel particles adsorbing on the fibrin filaments and modifying the structure of the semiflexible network. We develop two phenomenological models that quantify this hypothesis in physically distinct manners, and the derived predictions are qualitatively consistent with our experimental data.

**I. Introduction**

Stimuli responsive materials are important for diverse applications including actuators, sensors, tissue engineering, drug delivery, and soft robotics where they provide functionalities difficult to obtain by using passive materials [1–4]. Our interest here is in stimuli-responsive particle-filled polymer composites [5,6]. Such systems typically consist of a dense flexible chain polymer matrix (e.g., crosslinked elastomer) that exhibits a broad linear elastic regime and a mean physical mesh size that is small compared to the added rigid particles. Here we consider a different polymer-particle composite scenario based on large mesh semi-flexible biopolymer networks, which exhibit strong elastic nonlinearities, and particles that are soft and comparable in size to the network mesh.

From a property perspective, using a large mesh semi-flexible polymer network presents several opportunities for novel material design [7,8], especially biopolymer networks which are ubiquitous in living organisms. As background, we recall that biopolymer networks are present as both intracellular scaffolds and extracellular matrices which are vital for the structural integrity of biological structures [9–11], and play key roles in physiological processes such as cell migration, cytokinesis and mechano-responsiveness [12–15]. A key feature is that their mechanical properties are dependent not only on dynamic biochemical processes, but also on applied deformation. Biopolymer networks are well known for their distinctive nonlinear stiffening response under deformation (stress or strain), which distinguishes them from flexible polymeric networks (such as crosslinked elastomers) which generally show softening at large strains and overall are not nearly as deformation responsive [16–18]. In biological systems such a stiffening behavior is thought to protect tissues from large deformations thus preventing ruptures and maintaining tissue integrity [16,19]. The stiffness of some of these networks is also known to be controlled by molecular motors which can impose active stresses internally [15,20]. The nonlinear strain-stiffening behavior is a direct consequence of the stiffness and large persistence length of underlying polymer filaments which constitute these networks [21–25]. Such biopolymers are classified as semiflexible due to their large bending stiffness, and do not form random coils [24]. The semiflexible nature also strongly modifies the nature physical entanglements and influence of chemical crosslinking on mechanical response, resulting in structural and dynamical length scales which are very different from their flexible counterparts [26]. Large mesh networks have already found applications in tissue-engineering and cell culture scaffolds [27,28] because of such distinctive structural and dynamic-mechanical properties, and offer new mechanisms of developing stimuli responsive systems which are fundamentally different from the flexible polymer based composite systems.

Here, we report a two-component stimuli-responsive particle-polymer composite in an aqueous suspension which responds to increasing temperature by stiffening, even though the two components independently soften under the same conditions. We use as the matrix a large mesh semiflexible biopolymer fibrin network which exhibits a strain-stiffening deformation response. The biopolymers and corresponding network architecture is formed via in situ polymerization around a pre-existing dispersion of temperature-sensitive Poly(N-isopropylacrylamide) (pNIPAM)-based microgel particles. In dilute solution, pNIPAM particles collapse at temperatures above the lower-critical solution temperature (LCST) of ~32°C [29]. In the current work we employ a very large volume fraction of microgels (32-78%) in contrast to an earlier study [30] on a microgel-containing fibrin composite where the microgel volume fraction was less than 17% . Moreover, this earlier study did not focus on stimuli-responsive stiffening, but rather examined cell mobility through fibrin networks, showing that a composite of fibrin with microgel particles had enhanced cell mobility via the formation of colloidal tunnels compared to typical high density fibrin networks. Here, we show that at these large microgel volume fractions there is a near negligible modification of elastic stiffness of the composite network below the pNIPAM LCST where the microgels are swollen. Upon increasing temperature above the LCST, the composite controllably stiffens which we hypothesize is due to how microgel collapse modifies the fibrin network structure. This functional response was not present in the previous work on fibrin containing low volume fractions of microgels [30]. Furthermore, the mechanical behavior of our system is potentially richer at large microgel volume fractions since they can in principle form soft glass or gel-like states in the composite in a manner which could be modified by geometric or other fiber-induced physical factors. With growing efforts towards the development of synthetic semiflexible polymer networks [31–33], our efforts provide novel design principles for realizing new classes of smart responsive materials.

This remainder of this article is structured as follows. In Section II, we describe the experimental materials and methods. The characterization results for pure microgel particles, pure microgel suspensions, pure fibrin networks, and fibrin-microgel composites are presented in Section III. A discussion of the experimental results, along with the formulation and application of two distinct physical models, is presented in Section IV. The paper concludes in Section V. Additional details on materials synthesis, characterization, and modeling is provided in the Supplementary Information (SI).

## II. Materials and Methods

### II. A Microgel synthesis and characterization

Ultra-low crosslinked pNIPAM microgels were synthesized under a 'cross-linker free' condition following a published protocol [34]. Free-radical polymerization of NIPAM (Sigma-Aldrich) in water was initiated using potassium persulfate at 45°C and the temperature was gradually increased to 65°C in the absence of added cross-linker. The formation of microgel particles is attributed to self-crosslinking by chain transfer reaction during and after polymerization [35]. Typically, such microgels have a dense core and a more dilute/hairy corona structure which we assume to be, on average, a spherical soft object. The internal density $\rho(r)$ decreases continuously in a non-universal manner as one transitions from its center to edge.

As shown elsewhere [34], initiating polymerization at low temperature leads to a lower radical concentration in the solution and hence a low concentration of particle nuclei. Particle growth is achieved via favorable radical propagation relative to the nucleation of new particles process because of the higher concentration of monomer compared to nuclei. With a gradual increase in temperature more radicals are generated while monomer is consumed [34]. A detailed preparation protocol is provided in the SI section 1. A stock solution of 9 wt% polymer microgel particles was then diluted with Type 1 water to achieve the desired concentrations of the uncharged microgel suspension.

The particle radius as a function of temperature was determined by dynamic light scattering (DLS) (Zetasizer Nano ZS, Malvern) with a Helium-Neon gas laser emitting at 632 nm on a very dilute suspension (0.04 wt%) with a beam diameter of 0.63 mm. The temperature of the samples was controlled with a step size of 2.5°C by the automated heating/cooling module built in the instrument. The equilibration time at each temperature was 150 s.

**II. B Sample preparation**

The aqueous buffer used in all the samples was produced by mixing 25 mM HEPES, 150 mM NaCl, and 20 mM $CaCl_2$ in type 1 water and adjusting the pH to 7.4. Pure microgel suspensions were produced by diluting the stock solution of microgels with an estimated mass of aqueous buffer. Bovine fibrinogen (free of plasminogen and fibronectin) and thrombin were acquired commercially (Enzyme Research Laboratories, South Bend, IN, USA) and stocked at -70°C. Prior to composite preparation, fibrinogen was thawed at 37°C and diluted to the required concentration with the aqueous buffer. To prepare pure fibrin hydrogels, predetermined volumes of solutions of fibrinogen and thrombin were mixed in a centrifuge tube and immediately loaded between the plates on the rheometer. For preparing composites, a predetermined mass of pNIPAM microgels was added to the diluted fibrinogen solution in a microcentrifuge tube and the contents were mixed using a vortex mixer at 1000 r.p.m until the microgels were well dispersed visually. The solution was kept on ice and diluted thrombin was added to the fibrin-microgel solution to initiate gelation. The sample was quickly loaded between the rheometer plates and allowed to gel at a constant temperature of 25°C for an hour while probing its viscoelasticity using a very low oscillatory strain.

**II. C Rheological characterization**

Viscoelastic properties of all materials were probed using a torque controlled rotational rheometer (model Discovery Hybrid 3, TA instruments and MCR702, Anton Paar) using a plate-plate geometry. A 20 mm diameter steel plate was used in all experiments with a 600 grit, adhesive-back sand paper (Norton Abrasives) adhered to the contact surfaces to suppress wall slip. A fixed gap setting was used in each experiment. Typical gaps varied between 750 -1100 $\mu$m, significantly large compared to the microgel or polymer pore size, thus eliminating any confinement effects. The reported nonlinear measurements are without any parallel plate corrections. A solvent trap with a wet-tissue adhered to its interior was used to minimize water evaporation. The temperature of the bottom plate was controlled using a water-cooled Peltier system. For the pure microgel suspension experiments, all the samples were rejuvenated by shearing at 50 1/s for 60 s, and then relaxed for 12 min before taking data to suppress sample aging effects and erase any history [36]. For the pure fibrin and composite rheology, fibrinogen was allowed to polymerize for an hour while probing its viscoelasticity at an oscillatory strain with small amplitude $\gamma_0 = 1\%$ (in the linear regime) and an angular frequency $\omega = 1$ rad/s, after which time approximately constant plateau values of linear moduli were observed as a function of time.

Temperature sweeps from 25-35°C (up and down) were performed at a constant rate of 1°C/min while probing the linear viscoelasticity of the samples. Strain amplitude sweeps were also performed on the pure fibrin hydrogels and fibrin-microgel composites to probe the strain dependent viscoelasticity at a fixed angular frequency of 1 rad/s.

## II. D Microscopy

Two-color fluorescence microscopy experiments were performed on the composite using a laser scanning confocal microscope (LSM 700, Zeiss, Germany). A 63x oil-immersion objective lens (NA 1.4) was used. The microgels were labeled with rhodamine green (Invitrogen, USA) and fibrinogen was labeled with TAMRA-SE red (Thermofisher Scientific, USA) with excitation/emission wavelength of 504/535 nm and 547/576 nm respectively. The temperature of the sample was controlled using a heated stage (Delta T, Bioptechs, USA).

## III. Results

## III. A Microgel characterization

In dilute solution (0.04 wt%), the size of microgels strongly depends on temperature. Fig. 1 shows the average hydrodynamic diameter of the microgels measured via DLS. More details about the intensity distribution and polydispersity in size are provided in the SI Fig. S2. The averaged hydrodynamic diameter varies roughly from 2 to 0.6 µm as the temperature is varied between 25-35°C at 1°C/min with an abrupt decrease around the LCST of pNIPAM. This is attributed to an intramolecular "phase transition" of the pNIPAM polymer across the LCST and hence the volume of solvent the microgel can imbibe changes abruptly to achieve an equilibrium configuration [37,38]. Hysteresis is observed in the hydrodynamic diameter while cooling and heating the suspension which is consistent with other studies in the literature [39]. Microgels prepared by such a crosslinker-free preparation protocol are known to be highly deformable and ultra soft [34]. We did not explicitly measured the microgel stiffness, but our estimates using the Flory-Rehner theory [37,40] (see SI Sec 2) yields an elastic shear modulus of 1.6 kPa at 25°C and 4.6 kPa 35°C. This is consistent with our earlier work, where we have estimated a similar particle stiffness based on an independent statistical mechanical analysis [41].

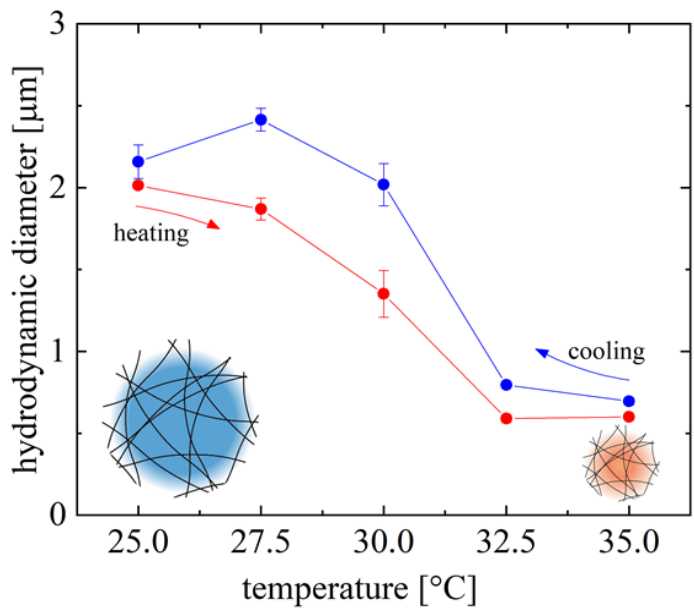

**Figure 1** Hydrodynamic diameter of self-crosslinked pure pNIPAM microgels, measured using dynamic light scattering, as a function of temperature. The "phase transition" in pNIPAM across the LCST leads to the sharp change in the volume of the microgel particles.

**III. B Pure microgel suspension rheology**

The linear elastic shear modulus $G'$ for pure microgel suspensions at various concentrations is shown in Fig. 2A as temperature is varied from 25 to 35°C. At all concentrations, the response is predominantly elastic ($G' > G''$) in the probed temperature range at 1 rad/s ($G''$ data is shown in SI Fig. S3). At low temperature, $G'$ monotonically increases with suspension concentration, owing to an increasing volume fraction of swollen microgels which are known to interact via repulsive pair potentials [41–43]. A linear concentration dependence ($G' \sim c_p$) is expected in the ultra-high volume fraction soft jamming regime and a stronger concentration dependence has been observed below it in the glassy fluid regime in earlier studies on similar microgel suspensions [36,41,44]. The power law dependence in the glassy fluid regime does not follow a universal scaling with concentration since it is dependent on microgel softness, chemistry, synthesis, etc. [36,41]. Based on the concentration dependence of $G'$ (Fig. 2B), we conclude that the explored concentration regime of pure microgels is in the glassy fluid regime, well below the soft jamming crossover. The origin of elasticity in such suspensions is associated with a kinetically arrested/caged state of microgels which do not relax on the observation time scales [41,45].

With increase in temperature (but below the LCST), the microgels undergo continuous and modest de-swelling, reducing the effective volume fraction of the suspension which results in a lower $G'$. Across the LCST (~32°C), the microgels shrink abruptly accompanied by the emergence of attractive interactions between the microgels which leads to the formation of local aggregates of de-swollen particles. We observe a non-monotonic trend in $G'$ versus concentration at temperatures close to the LCST. This may be attributable to the inhomogeneity in the suspension that has low effective volume fraction of aggregated de-swollen microgels. The repeat measurements are shown in SI Fig. S4. A key point to note is that $G'$ drops by nearly an order of magnitude as the temperature is increased from 25-35°C in the glassy fluid regime. Unlike previously published literature on pNIPAM based microgel suspensions, which stiffen across the LCST due to attractive interactions between the microgels [46], we observe no re-emergent stiffening in $G'$ above LCST. We attribute this lack of stiffening to the lower concentrations (well below soft jamming) and the different chemistry of the microgels used in the current work.

A major challenge in working with suspensions of soft microgels is determining the effective volume fraction. The compressible microgels can swell/de-swell depending on osmotic pressure which is as a function of concentration. We do not have *a priori* information about the size of microgels as a function of varying concentration or osmotic pressure. Hence, to estimate the volume fraction of microgel suspensions at low temperature, we adopt an approach suggested by recent experiments on similar microgels [47]. In the dilute low concentration regime, the microgel size is fixed and one can define an effective (hydrodynamic) volume fraction by determining an intrinsic viscosity [$\eta$] from the steady shear experiments with very dilute suspensions (SI Fig. S9). From this, we obtain a volume fraction $\phi = 0.22$ at $c_p = 0.25$ wt%. In an intermediate concentration regime of $c_{p1} < c_p < c_{p2}$ per SI Fig. S10, the microgels begin to de-swell due to steric repulsions between particles, in a manner that experiments suggest is initially weak. Crudely, experimental data (of microgel radius $R$) in the latter regime can be modeled as a power law of $R \sim c_p^{-1/6}$, implying an effective volume fraction that scales as $\phi \sim c_p^{1/2}$. Beyond some "high enough" $c_p > c_{p2}$, one expects that the more fuzzy "corona" part of the microgel is largely squeezed out, leaving a dense core which at yet higher concentration undergoes isotropic compression in the sense that $R \sim c_p^{-1/3}$, as again suggested in literature [47]. This leads to $\phi \sim c_p^0$ where the linear growth in microgel concentration is perfectly compensated by their shrinking size. Ultimately beyond an even higher concentration $c_{p3}$, the internal concentration of microgels presumably saturates at a maximum value similar to that of a collapsed globule with radius $R_{\text{collapsed}}$. In our experiments we always remain below $c_2$, i.e. in the glassy fluid regime. Hence, for the range of concentration 0.3-3wt% that we have used in preparing composites, the volume fraction at 25°C varies from $\phi = 0.28$-$0.78$ as shown in Fig. 2B. We can also estimate an effective volume fraction (a lower bound) at temperatures beyond the LCST assuming all the microgels shrink to the same collapsed size as indicated by the DLS experiments. The volume fraction in the deswollen state 35°C varies from $\phi = 0.012$-$0.072$. We will use the obtained values of effective volume fraction (SI table S6) in Section IV.

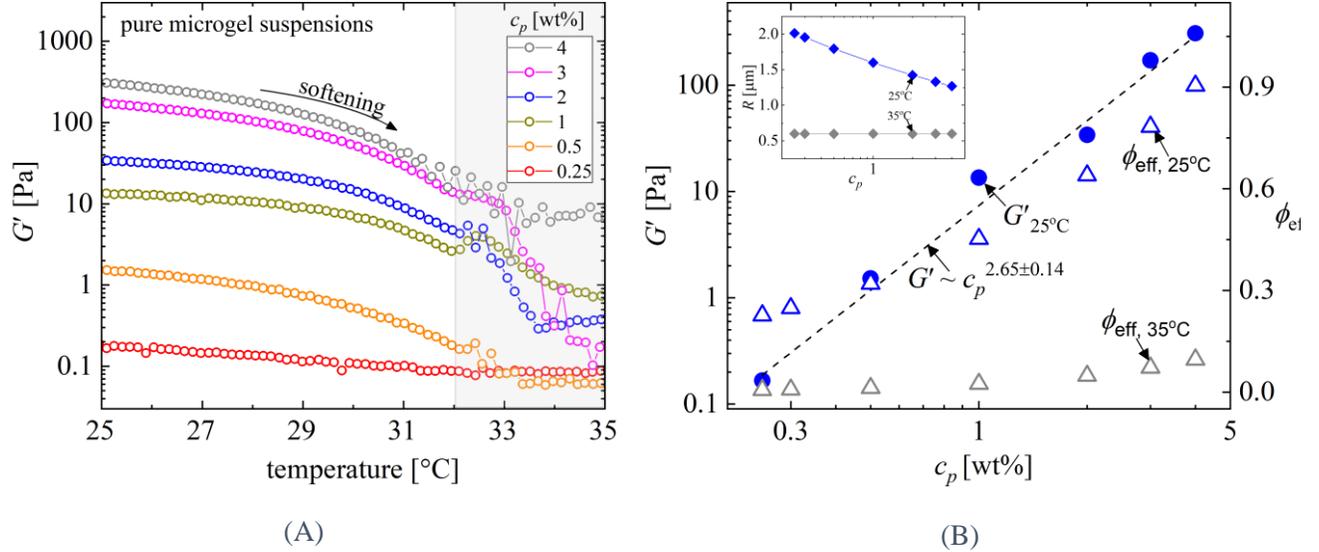

(A) (B)

**Figure 2** Pure pNIPAM microgel suspension shear elastic modulus $G'$ as a function of temperature and concentration. (A) At all concentrations, $G'$ decreases with temperature. $G'$ values above the LCST (shaded region) show a non-monotonic trend with concentration. This is attributed to the sample inhomogeneity above the LCST (repeat measurements in SI Fig. S4) (B) Concentration dependence of $G'$ at low temperature, 25°C, shows a strong apparent power-law (best fit shown by the dashed black line) which is indicative of concentrations in the glassy fluid regime before the soft jamming crossover. For context, the estimated effective volume fraction $\phi$ and particle radius (evaluated as described in Sec. III B) is shown for both low and high temperature (see right vertical axis).

### III. C Pure fibrin hydrogel rheology

Fibrinogen polymerizes into a polymer network of fibrin filaments in the presence of thrombin following a cascade of events [48]. Pure fibrin networks have a viscoelastic solid-like response in the linear response regime. The concentration dependence of the linear elastic shear modulus $G_0$ (Fig. 3A) follows a power-law dependence on the concentration of fibrinogen; here we observe $G_0 \sim c_f^{2.1\pm0.1}$. The power-law exponent is close to the theoretical scaling exponent of 2.2 for these entangled semiflexible polymer networks at low-strains due to an entropic single fiber stress storage mechanism [21,49]. Fig. 3B shows the nonlinear shear strain-stiffening response of pure fibrin networks at various concentrations (at 25 °C) when probed with an oscillatory strain of increasing strain amplitude $\gamma_0$. The stiffening, indicated by the increasing first-harmonic (average) elastic modulus $G_1'$, is most apparent when the imposed shear strain is greater than 10-20%. This is attributable to the nonlinear force displacement relationship of the individual strands as their thermal slack is reduced with external deformation [21,25,49]. Alternative explanations for such strain stiffening also exist in the literature based on the nonaffine network rearrangements in the network [50]. The average shear elastic modulus $G_1'$ at the maximum strain before the network breaks irreversibly is roughly 10 times higher than the linear elastic modulus $G_0$. As a function of increasing temperature, the pure fibrin networks show a very weak softening at all concentrations used here (Fig. 5, SI Fig. S7).

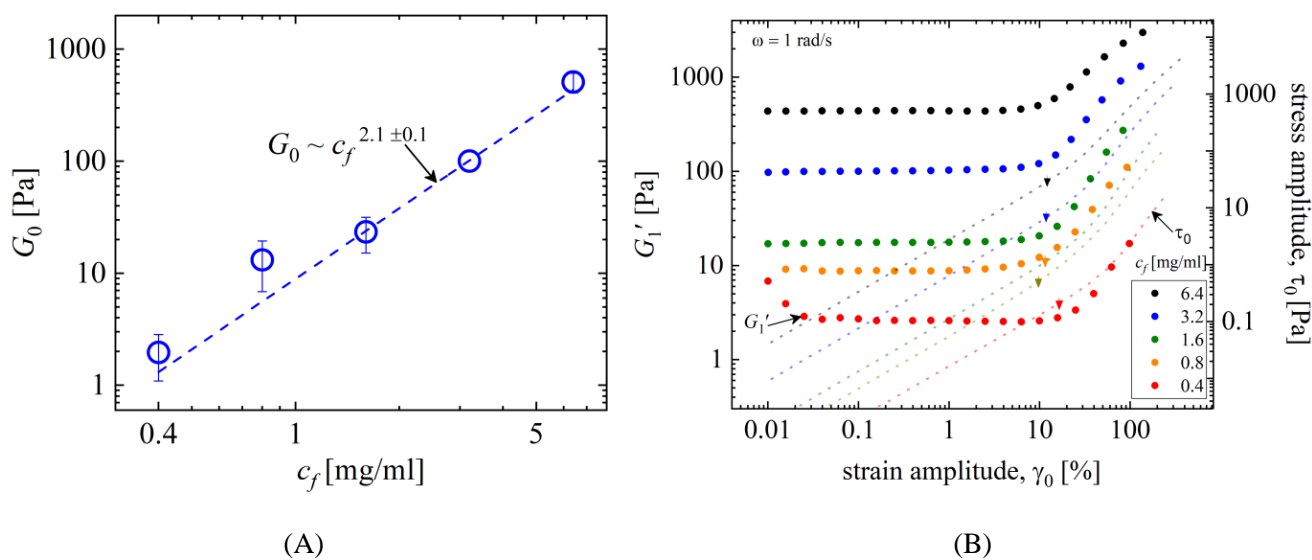

(A)            (B)

**Figure 3** Pure fibrin rheology as a function of concentration and shear strain amplitude. (A) The concentration dependence of the linear elastic modulus $G_0$ agrees closely with the theoretical predictions for an entangled semiflexible polymer networks. (B) Strain stiffening response of the oscillatory first-harmonic elastic modulus $G'_1$ (filled symbols) is prominent when shear strains exceed 10-20%. Similar effects can be visualized in a stress amplitude $\tau_0$ vs strain amplitude $\gamma_0$ plot (dotted lines), where the nonlinear effects are evident when the stress-strain relation deviates from linearity as indicated by the arrowheads. The normal force measurements for these experiments is provided in SI Fig. S5.

**III. D Fibrin-microgel composite rheology**

Figure 4 demonstrates that fibrin networks readily form in the presence of these microgel colloidal particles at the concentrations studied (see SI Fig. S6 for all other compositions). Polymerizing fibrinogen in the presence of colloidal particles is a complex process since colloids can disrupt network formation [8]. We observe similar formation timescales even in the presence of microgel particles and the linear moduli approach a constant plateau at long times, indicating that all the available fibrinogen has been converted into fibrin. Composites are predominantly elastic ($G' > G''$) with an elastic modulus similar to that of pure fibrin (c.f. Fig. 3A).

The particles do tend to slightly increase the elastic modulus of the composite, though the trend is non-monotonic and seemingly unsystematic. It is striking that the effects are so small, considering the high effective volume fraction $\phi$ of the microgel particles (from Fig. 2B, for $c_p \geq 1$wt%, $\phi \geq 45\%$,). In traditional polymer-colloid composites involving 'hard' colloidal inclusions in a flexible polymer network, the mechanical stiffness is more dramatically increased by the reinforcement effects from inclusions [51]. We attribute the weaker inclusion effects here to several distinct features of the fibrin composites. Foremost is the fact that fibrin networks have large pore sizes compared to the inclusion, i.e. pore size is on order of microns, similar to the diameter of the microgel particles, unlike in flexible polymer composites where the mesh size is typically much smaller than the inclusion. Additionally, the preparation protocol involves gelling a premixed solution of fibrinogen and microgels. We expect some fraction of fibrinogen may get adsorbed on the particles and becomes unavailable to form the fibrin network as indicated in one previous work [8]. Furthermore, the salt concentration in the hydrogel composite will vary as a function of the microgel concentration since the microgel stock solution was prepared with deionized water. The architecture and mechanics of fibrin networks are known to depend on salt concentration [48]. Finally, the microgels are soft and very lightly crosslinked, hence there can be a possible modification of microgel coronas due to the fibrin network [30] and fibrin polymer chains may penetrate the microgel corona. Any further investigation into such very complex effects is beyond the scope of this work, but it is clear from Fig. 4 and Fig. S6 that composites successfully form and at fixed temperature, the microgel inclusions have only a slight effect on the elastic modulus.

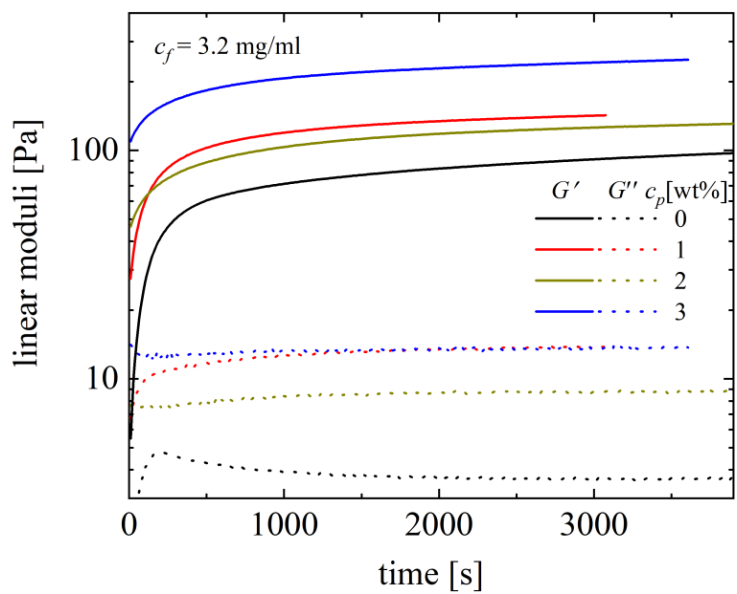

**Figure 4** Composite formation by polymerizing fibrinogen into a fibrin network in the presence of pNIPAM microgel suspensions at different particle loading $c_p$ [wt%]. Evolution of viscoelastic moduli show weak effects from the pNIPAM that are non-monotonic but tend to increase the overall linear moduli. (see SI Fig. S6 for all other compositions)

Increasing the temperature shrinks the microgel particles, decreasing their effective volume fraction. But in qualitative contrast to standard composites, this results in a striking *increase* of the overall elastic stiffness of the composite. Figure 5 shows the linear viscoelastic moduli and how they respond to temperature cycling. Two extreme compositions are shown in Fig. 5, a high concentration of both fibrin and microgel ($c_f$ = 6.4 mg/ml, $c_p$ = 3wt% microgels) and a low concentration of both fibrin and microgel ($c_f$ = 0.4 mg/ml, $c_p$ = 0.5wt% microgels). Thrombin concentration is held fixed at 0.5 U/ml for both the composites. Although the fibrin concentration (and hence the elasticity) is different in both the cases, the temperature-dependent response is very similar. As the temperature is increased at a rate of 1°C/min, initially at 25-28°C a weak decrease in the linear shear moduli ($G'$, $G''$) is observed, followed by a sharp increase in the linear moduli within a narrow temperature range of 28-32°C around the microgel LCST. Further increase in temperature (32-35°C) leads to a weak increase in the moduli.

The response is reversible in the cooling cycle accompanied by a hysteresis. The latter can be attributed to hysteresis in the deswelling/swelling of pure microgels during heating/cooling cycles (Fig. 1).

The temperature-dependent response of the composite is contrary to that observed for pure microgel suspensions (Fig. 2A) and pure fibrin network (SI Fig. S7), where $G'$ monotonically decreases with increasing temperature. The normal force $F_N$ variation during the heating/cooling cycles is also shown in Fig. 5. No clear trend for a composite with 0.4 mg/ml fibrin is seen, with $F_N$ roughly constant, i.e. changes are less than the instrument resolution of 0.05 N (i.e. a resolution of 300 Pa average normal stress $\sigma_N$ on the bounding surface, $\sigma_N \approx 2F_N/\pi r^2$, $r$ being the radius of the sample/plate). However, for the higher-concentration composite ($c_f = 6.4$ mg/ml) we observe a measurable $F_N$ during the heating and cooling cycles. The normal force $F_N$ acts to pull down the upper rheometer plate (gap is kept fixed) during the heating cycle as temperature crosses the LCST, up to $F_N = -0.11$ N (700 Pa), while during the cooling cycle this normal force vanishes. This is likely a consequence of decreasing volume of microgels as they shrink, resulting in a tensile force within the network and a downward force on the upper plate.

Beyond the LCST, temperature-induced stiffening is observed for all compositions studied. Fig. 6 shows the effect of increasing microgel concentration for the two extremes of high and low fibrin concentration (data for all compositions is given in SI Figs. S7 and S8). In Fig. 6 and SI Fig. S7-8, only the second heating cycle is shown for comparison. Increasing the concentration of microgel particles increases the relative stiffening of the linear moduli. We also observe a similar response in $G''$ but for all cases the response is predominantly elastic at all temperatures, i.e. $G' > G''$. Similar observations were made for other fibrin concentrations ($c_f = 0.8, 1.6, 3.2$ mg/ml) (SI Fig. S7-8). The mild softening of the modulus in the range 25-28°C appears to follow the same trend as the softening of the shear modulus of pure microgel suspensions. However, this effect is not noticeable when the modulus of pure fibrin is significantly higher than that of pure microgel suspensions. The weak stiffening of the modulus in the range of 32-35°C is directly correlated to the weak decrease in the microgel size.

To compare the thermoresponsive stiffening across the concentrations, we plot in Fig. 7 the stiffening observed in all cases as the dimensionless ratio, $G'_{35°C}/G'_{min}$. Here, $G'_{min}$ is the minimum value of the composite elastic modulus before it begins to stiffen (indicated in Fig. 6B), and $G'_{35°C}$ is the modulus at 35°C. Some general trends are evident from Fig. 7. A higher concentration of microgel particles leads to a higher stiffening ratio. The stiffening in terms of absolute difference is even more dramatic, as shown in the inset. The fibrin concentration also plays a significant role. The absolute changes (Fig. 7 inset) are larger at higher $c_f$, i.e. for stiffer networks. But, interestingly, the relative changes (main Fig. 7) are larger at lower $c_f$, i.e. softer networks have more dramatic relative changes in stiffness. A wide range of baseline modulus is covered by these compositions since the linear modulus of the fibrin network scales sensitively with concentration ($G_0 \sim c_f^{2.1}$, Fig. 3A).

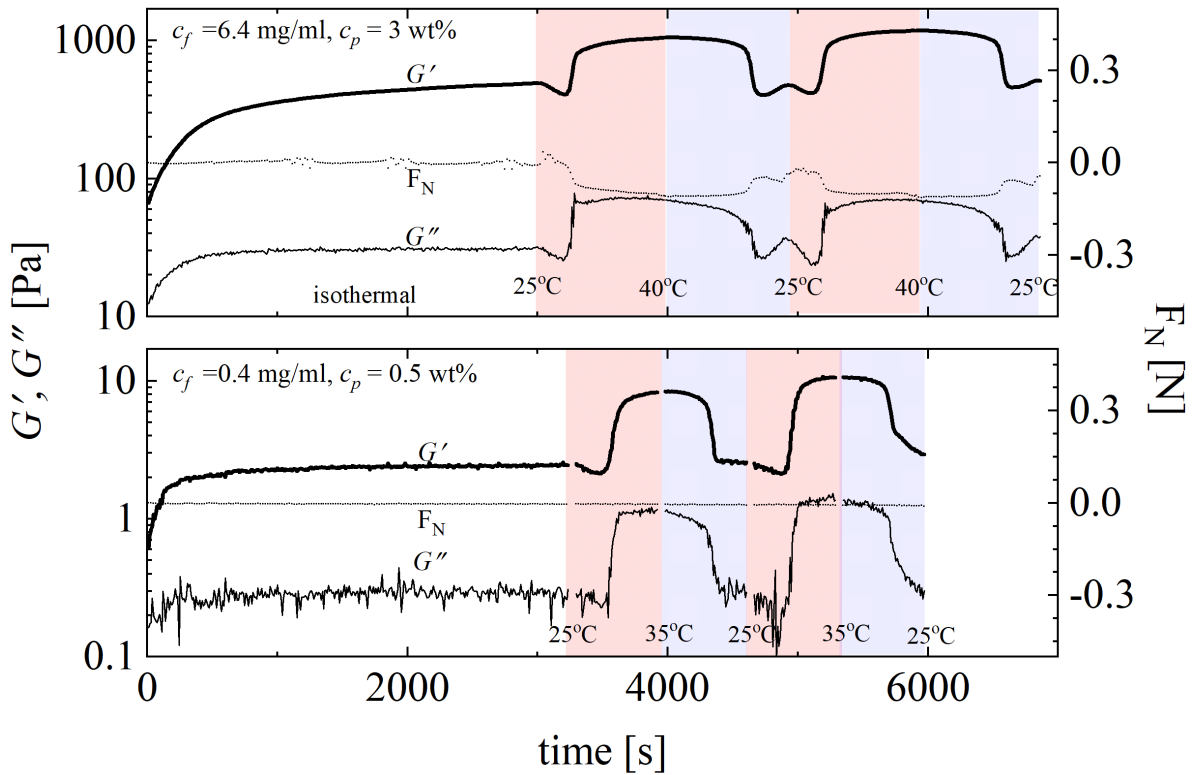

**Figure 5** Dependence of composite rheology on temperature. The linear moduli $(G', G'')$ and normal force $F_N$ for two different compositions of composites: 6.4 mg/ml fibrin with 3wt% microgels (top) and 0.4 mg/ml fibrin with 0.5wt% microgels (bottom) are shown. The composite is prepared by polymerizing fibrin in the microgel suspension at a fixed temperature 25°C followed by the application of repeated temperature increase (shaded red) and decrease (shaded blue) at 1°C/min to probe the temperature dependence. The linear moduli show reversible stiffening as the temperature is raised above the LCST (strain amplitude 1%, angular frequency 1 rad/s).

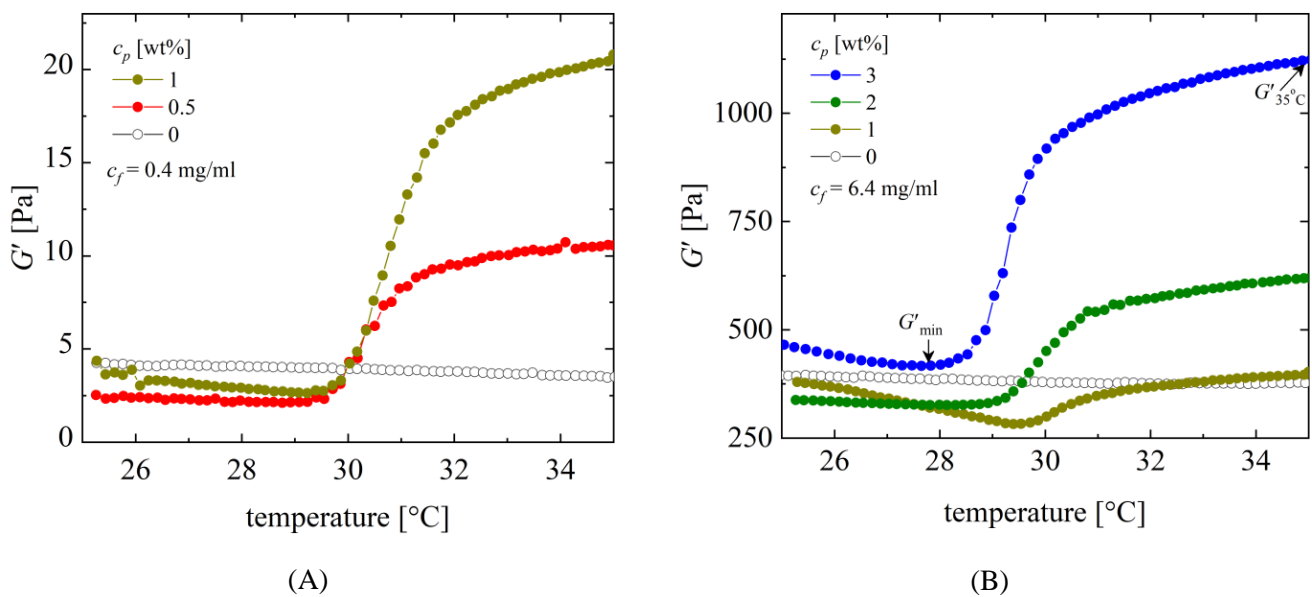

**Figure 6** Temperature dependence of the storage modulus of pNIPAM-fibrin composites. (A) The concentration of fibrinogen and thrombin is held constant at 0.4 mg/ml and 0.5 U/ml, respectively, while the concentration of pNIPAM microgels is varied. Varying the concentration of pNIPAM microgels stiffens the network as the temperature is increased above LCST. (B) Similar data for 6.4 mg/ml fibrin and 0.5 U/ml thrombin. Note that only the second temperature ramp-up cycle is shown in the figures. Data for all other compositions is given in SI Fig. S7, and repeat measurements at selected compositions are shown in SI Fig. S8.

We used confocal imaging to investigate the structure of the fibrin matrix in the composites and test our assumption that microgel particles are shrinking within the network at elevated temperature. Fig. 8 shows the confocal images of the composite at 25°C and 35°C. Qualitatively, it appears that the presence of microgels does not interfere with the formation of the fibrous network, which appears similar to that of pure fibrin hydrogels. The volume fraction of microgels decreases as the sample is heated from 25 to 35°C as is apparent by the decreasing intensity of emission from tagged microgel particles, consistent with microgel particle shrinking at elevated temperatures. However, no apparent change can be observed in the fibrous network visually between 25 to 35°C. We do not make a quantitative comparisons of the network topology features with and without the microgels.

Fig. 9 verifies that strain-stiffening is retained in the composites, a property native to pure fibrin (Fig. 3B) which, as we hypothesize in section IV, may play a central role in the temperature-induced stiffening caused by de-swelling microgel particles. Large-amplitude oscillatory shear (LAOS) results in Fig. 9 are shown in terms of the first-harmonic elastic modulus $G_1'$ (representing the average energy storage during the deformation [52]), as the shear strain amplitude $\gamma_0$ is varied at 25 and 35°C for three different compositions of the composite. At 25°C, all three composites show strain-stiffening behavior similar to pure fibrin (Fig. 3B). At 35°C, the composites still show strain stiffening, but it is less dramatic. The reason is associated with a higher initial plateau modulus in the linear regime, but a similar final elastic modulus at the largest imposed strain. The retention of strain-stiffening features in the fibrin matrix below and above the LCST suggest that the large deformation rheological response in the composite is being dominated by fibrin, i.e. the presence of microgels during the fibrin polymerization may not affect the underlying mechanism of strain-stiffening of the network. However, as mentioned earlier, the salt concentration in the composites is different than the pure fibrin network, which may lead to architectural/topological differences between the composite fibrin networks and pure fibrin networks [48].

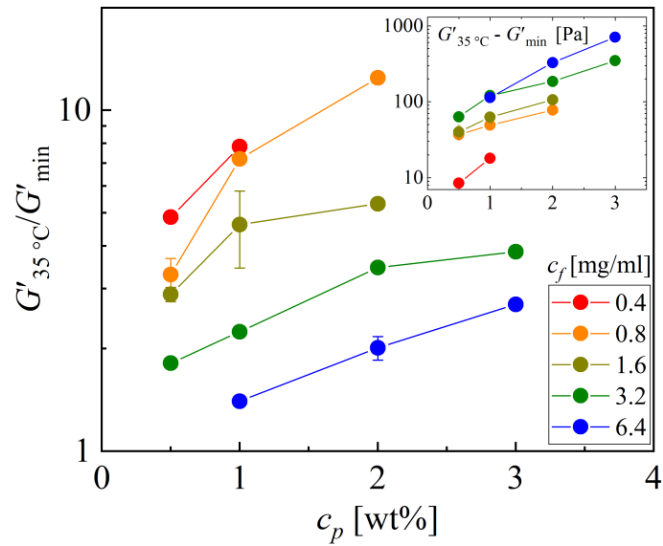

**Figure 7** Stiffening ratio for all fibrin concentrations $c_f$ plotted as a function of microgel content $c_p$ in the composites. (Raw data for $c_f$ = 0.4, 6.4 mg/ml in Fig. 6; all other raw data in SI Fig. S6-7.)

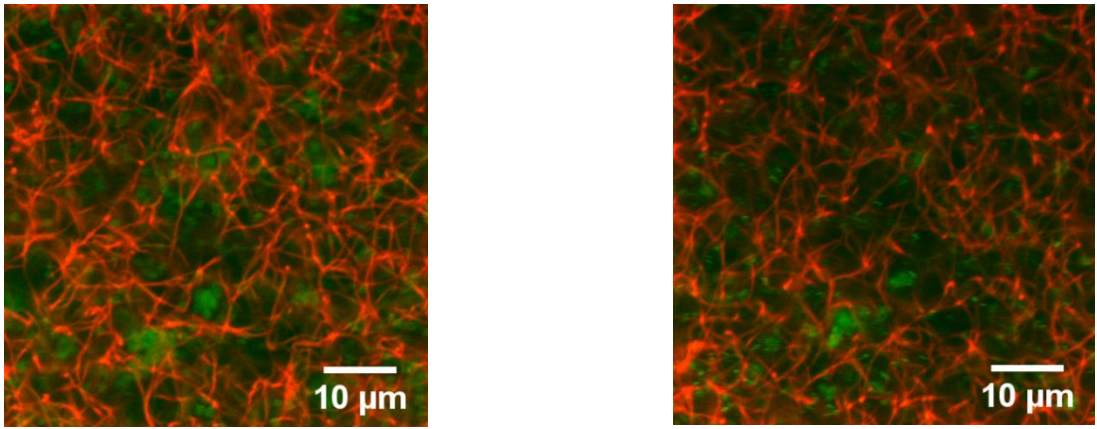

(A) T = 25°C  (B) T = 35°C

**Figure 8** Confocal images of fibrin-microgel composite at (A) T = 25°C and (B) T = 35°C. The fibrin fibers (tagged red) and the microgels (tagged green) have a concentration of $c_f$ =3.5 mg/ml and $c_p$ = 3 wt%, respectively. Above the LCST(~ 32°C), de-swelling of microgels can be observed.

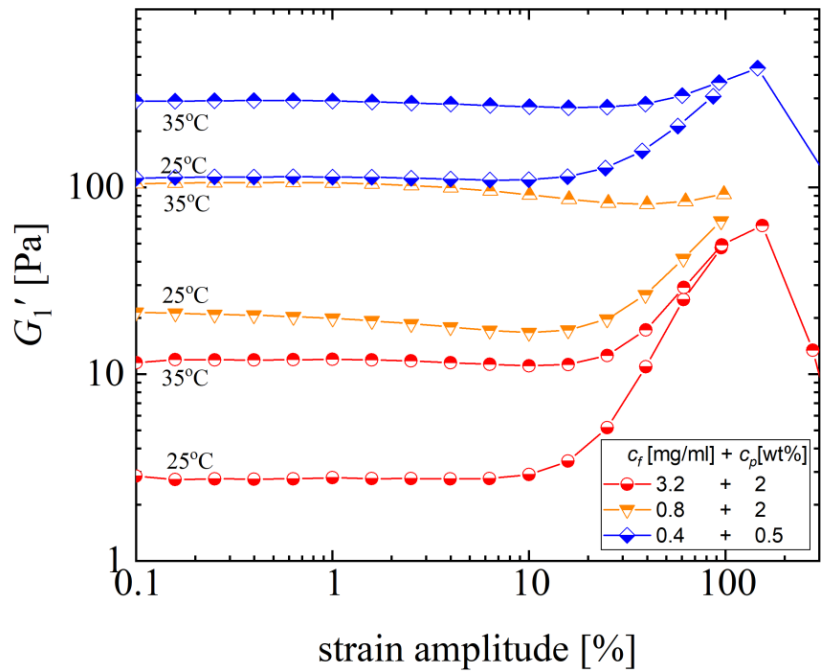

**Figure 9** Nonlinear response of fibrin-microgel composites at selected compositions, each at low and high temperature (compared to LCST). Both below and above the LCST, the composites retain a strain-stiffening response similar to pure fibrin hydrogel in Fig. 3 B.

**IV. Discussion and Theoretical Modeling**

Although soft microgels have been previously incorporated into fibrous matrices to modulate properties (e.g. transport [30] and mechanical [53–55]), the thermoresponsive sensitivity we report here is unique. While the individual components mechanically soften when the temperature is raised from 25-35°C, the composites show significant stiffening. The fundamental question then arises as to the physical mechanism. There are various speculative hypotheses in the literature that might be relevant to our observations, although the problem is very complicated for many reasons including the crucial aspect that the precise nature of the coupling between the components is not well understood. A prior study of pNIPAM-based microgels (fabricated by a different approach) embedded in a flexible polyacrylamide (PAAm) matrix found a modest enhancement (2- 4 times the matrix modulus) of material stiffness above the LCST [55] which was attributed to a transition of the microgels from soft fillers to hard fillers. A macroscopic continuum mechanics model was invoked as the origin of an increased composite elastic modulus. Various empirical relations have been proposed over the years to predict reinforcement [51] based on this perspective. The most common is a classic rigid particle inclusion mechanism [51], analogous to Einstein's equation for effective shear viscosity, which for the shear modulus is known as the Gold-Guth relation: $G_c/G_m = 1 + 2.5\phi_P$, where $G_c$ is the composite shear modulus, $G_m$ the pure polymer matrix shear modulus, and $\phi_P$ is the volume fraction of rigid filler sphere assumed to be perfectly adhered to the continuous polymer matrix. Of course, such a model is physically inappropriate for our system since the microgels are relatively soft and even in their de-swollen state are not large compared to the mechanical length scale (polymer mesh) of the matrix. Nevertheless, if we adopt this perspective, then based on our volume fraction estimates for the range of microgel concentration studied (above the LCST, only $\phi = 0.012 - 0.07$) one predicts $G_c/G_m$ is only ~ (1.03-1.17), far smaller than we observe. We therefore conclude the stiffening mechanism observed in our experiments is qualitatively different from that proposed for previously studied systems including [55].

Our limited goal in this section is to formulate speculative, but we think plausible, microscopic physical models for the stiffening mechanism. Two mechanisms are considered, both based on the idea of microgel-induced modification of the fibrin network beyond the LCST. We hypothesize microgels de-swell at high temperatures in the composite as they do in dilute solution, and a strong fibrin-microgel attraction (of hydrophobic or other physicochemical origin which does not break on the relevant experimental time scale) allows the microgel shrinking to pull on the surrounding network. This induces a structural change in the fibrous mesh surrounding the de-swollen microgels and results in an overall stiffening of the composite.

The problem, in principle, could be approached from an equilibrium thermodynamics perspective where the total composite free energy is minimized. The latter involves the competing effects of distortion of the fibrous mesh, polymer-microgel interactions, and microgel deswelling. But to quantify these factors requires knowledge of the details of connectivity and interactions between the fibrous mesh and microgel, the nature of junctions in the fibrous mesh (chemical crosslinks or physical entanglements), and free energy functionals for the fibrin network and microgels. Such an analysis is beyond the scope of current work. Rather we construct two simpler physical models which differ with regard to how the fibrin network is hypothesized to change when microgels shrink: (i) an effectively stretched polymer network model (fixed junctions), and (ii) an adjusted mesh model (mobile junctions). In both cases, the deswelling of microgel particles will generate contractile forces in the fibrous network to which they are attached. However, the effects of such contractile force is realized differently in the two models, and thus the underlying physical origins of stiffening are different. Based on the current experimental evidence, we do not have a definitive way of proving/disproving either model, and it is likely the real physical behavior contains aspects of both conceptual pictures.

**IV.A. Stretched network model (fixed junctions)**

The observation of a negative pull (normal force) on the upper rheometer plate above the LCST indicates that a net tensile stress exists in this state. We hypothesize that microgel deswelling results in a microscopic stress in the fibers which manifests itself as a bulk contractile pull in the network. Such microscopic deformations may be responsible for the stiffened composite response above the LCST. For example, the average value of the tensile (normal) stress $\tau_e \approx 700$ Pa for the sample with $c_f = 6.4$ mg/ml and $c_P = 3$ wt% (from Fig. 5) is significantly higher than the average shear stress at the onset of nonlinear strain stiffening of the network ($\tau_c \approx 50$ Pa, obtained by $\tau_c \approx$ $G_0 \gamma_c$, where $\gamma_c$ is the characteristic strain at which the shear response begins to show measurable nonlinearity, in this case $\gamma_c \approx 0.1$ from Fig. 3B). However, a direct comparison cannot be made since these measured normal stresses arise from different modes of deformation. Further, the normal stresses at other composite compositions were below the instrument measuring limit, and hence cannot be reliably resolved.

We note that the hypothesis of microscopic internal strain induced stiffening was invoked in recent work involving multiaxial deformation of fibrin networks [56]. These workers hypothesized that external uniaxial extension applied to the network removes undulations in some of the fibers (along with changing the orientation of fibers from the undeformed state) and results in a higher shear modulus, whereas uniaxial compression induces a higher level of undulations and buckling, resulting in a softening response of the network. Deformation fields in that work are qualitatively different from our experiments. In the case of our fibrin-microgel composite, the deformation fields are spatially inhomogeneous on the scale of the mesh surrounding the microgel particles. In the simplistic scenario of a fibrous mesh bound to a deswelling microgel, we expect the fibers aligned in the radial direction experience a stretch and fibers aligned orthogonal experience a compression. The bulk network response will thus have two contrasting contributions: a stiffening response from the stretched fibers and a softening response from the compressed fibers. From our experimental observations, the mechanical response from stiffened fibers dominates the bulk response. The nonlinear shear response of our fibrin-microgel composites (Fig. 9) shows similarities to fibrin networks with imposed uniaxial extension [56]. In the pre-stressed state (here above LCST, in the prior work for extensional pre-strain), the onset of stiffening shifts to a higher value of applied shear strain in all the cases probed. Thus, we suspect similar underlying origins of the induced stiffening, i.e. the nonlinear deformation (strain-stiffening) response of individual fibers.

To test if such a hypothesis is physically consistent with our experimental observations, we adopt a constitutive model for semiflexible polymer networks from the literature. The polymer network is assumed to have *fixed* junctions (e.g., effective chemically crosslinked at the branch points or nodes). The Euclidean distance between the nodes, $l$, is different than the fiber arc length, $l_{arc}$, such that the normalized end-to-end distance, $x = l/l_{arc} < 1$ in the reference state. Adopting a mean-field-like approach to model the fiber inextensibility, an analytical form for the free energy of a semiflexible polymer chain as a function $x$, has been previously derived [23]

$$u_{fiber} = k_B T \pi^2 \nu (1-x^2) + \frac{k_B T}{\pi \nu (1-x^2)} \tag{1}$$

where $k_B$ is Boltzmann's constant, $\nu = \kappa/2k_B T L_c$ is a dimensionless stiffness parameter with bending rigidity $\kappa$ defined as $\kappa = l_p k_B T$, and $l_p = 32$ μm is the fiber persistence length assuming bundles of fibrin filaments exist [18] (a precise value of $l_p$ depends on the fibrin concentration amongst several other factors but for the purpose of simplicity of analysis, we assume a fixed value at all concentrations). The two terms on the right hand side of the Eq. (1) represent the contributions from fiber bending energy and entropy, respectively. The force-displacement ($f$-$x$) relationship for such a fiber derived from **Error! Reference source not found.** diverges as $x \to 1$, $f \sim (1-x)^{-2}$, as a result of finite extensibility of a fiber pulled taut when $l = l_{arc}$. Equation (1) has been previously employed [23,25,57] to constitutively model the nonlinear shear elastic response of semiflexible polymer networks assuming a 3-chain and an 8-chain network model. The advantage of using cubic lattice models is the ability to obtain analytic expressions for the shear stress. We use a 3-chain model since it captures some salient features of semiflexible polymer networks such as negative normal stresses in shear in addition to generic strain-stiffening effects. We refer the readers to previously published work for details of the models [25,57].

We employ the constitutive relation (Eq.7 of [25]) for the nominal linear shear modulus $G$ in the limit of zero shear strain for a 3-chain network model given by

$$G(x,\nu) = \frac{2}{3} n_f k_B T x^2 \left[ \frac{1-x^4}{\nu \pi [1-2x^2+x^4]^2} - \nu \pi^2 \right] \tag{2}$$

where $n_f$ is the density of crosslinked chains. Equation (2) was derived for a pure semiflexible polymer network in which each mesh/lattice cell is deformed affinely with an external shear strain. The deformation field in the network resulting from a deswelling microgel is more complex and highly dependent on the geometry and nature of fiber-microgel contacts. A quantification of such inhomogeneous deformation of the mesh is possible in principle but is beyond the scope of the current work.

To simplify the analysis, we therefore adopt a mean field like approach. Based on our experimental observations of a net stiffening response, we assume that microgel particle deswelling results in the entire fibrous network changing to an "effectively stretched" state. In the stretched state some of the thermal undulations in the fibers are removed compared to their conformations below the LCST, i.e. $x > x_0$, where $x_0$ is the fiber normalized end-to-end distance for the reference state ($25^\circ\text{C}$). As a result we expect the individual fiber response to stiffen, which then results in bulk network stiffening. Our approach assumes that the effective stretch in the entire network is uniform regardless of the spatial orientation and location of fibers relative to the microgel. The effective stretch $\lambda$ is defined from the change in the volume of the fibrous network compared to a reference state.

$$\lambda^3 = \frac{V}{V_0}. \tag{3}$$

We will consider the fibrin network at 25°C (initial state) and 35°C (final state). The volume of the fibrous network is the fraction of the total volume of the composite $V_t$, based on the theoretical effective volume fraction $\phi_p$ of microgels, $V = (1-\phi_p)V_t$.

The theoretical microgel de-swelling may not exactly transfer to a volume increase of the fibrin phase due to, for example, imperfect adhesive contact between microgels and fibrin, or a good adhesive contact for which the stiffness of the fibers resists microgel de-swelling. We therefore introduce a prefactor $m$ to map the theoretical to the actual volume increase of the fibrin phase. For the case of perfect binding and complete de-swelling, $m=1$, whereas the other extreme $m=0$ applies if the fibrin network volume is unchanged. For realistic scenarios we expect $0 < m < 1$. Hence, we can write

$$\lambda_f = m\lambda = m\left(\frac{1-\phi_{p,35^\circ C}}{1-\phi_{p,25^\circ C}}\right)^{1/3} \tag{4}$$

where $\phi_{p,35°C}$ and $\phi_{p,25°C}$ are the theoretical effective microgel volume fractions at 35°C and 25°C respectively. Since the free energy contribution from fibrin is highly nonlinear Eq. (1) and strongly depends on $\lambda$, it is reasonable to expect beyond a certain value of $\lambda$ that the free energy cost of deforming the fibrin network becomes larger than the lowering of free energy associated with full deswelling of the microgels. Hence, in equilibrium the microgel deswelling may be limited by the deformation of attached fibrin matrix and the particles would not achieve exactly the same size reduction as in dilute suspension, and thus $\lambda_f < \lambda$.

Within the mean field approach, the fibers are in a pre-stretched state when the microgels shrink with end-to-end distance $x = \lambda_f x_0$. The shear modulus is predicted to change, based on Eq. (2), from $G(x_0)$ to $G(x)$. The relative stiffening as the temperature is varied across the LCST can now can be expressed as the ratio $\Sigma = G(x)/G(x_0)$. Being more explicit about functional dependence, $\Sigma = \Sigma(x, x_0, \nu)$ where $x_0$ and $\nu$ are set by the fibrin concentration, $c_f$. The stretched state $x = \lambda_f x_0$ depends on the factor $m$ and the microgel effective volume fractions $\phi_P$ through Eq.(4), and $\phi_P$ in turn depends on the microgel concentration $c_p$. Thus the stiffening ratio can be expressed in terms of known material composition $(c_f, c_p)$ and the unknown factor $m$ as

$$\Sigma(c_p, c_f, m) = \frac{G(m; c_p, c_f)}{G_0(c_p, c_f)} \quad . \tag{5}$$

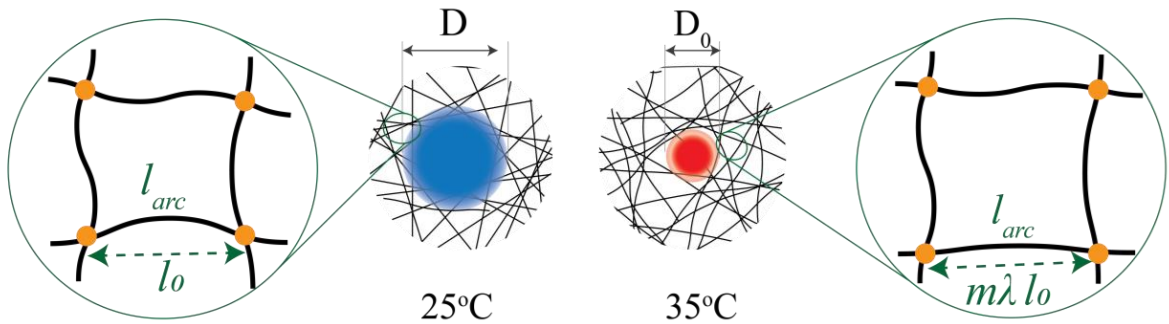

Figure 12 **Effectively stretched polymer network model.** At low temperature (T < LCST), the fibrin network is polymerized around the swollen microgel. The sketch of a single two dimensional lattice shows the key length scales in the network: the contour length $l_{arc}$ of the polymer in between the nodes/crosslinks, and the end-to-end

distance $l_0$ between the crosslinks. When the temperature is raised above the LCST of the microgel, its size abruptly shrinks to $D_0$ and this imposes a strain on the attached fibrin network. The consequence of this strain is in a mean-field-like sense is schematically shown in the two dimensional lattice sketch, where the end-to-end distance changes to $\lambda_f l_0$ but $l_{arc}$ remains constant as a result of the inextensibility assumption.

In Eq. 5, $m$ is the only unknown fit parameter. Other parameters are a priori known from the fibrin and microgel concentrations, and we use literature data to determine values of $l_0$ and $l_c$ (see SI table S1). Figure 13 A shows the computed value of $\Sigma$ as a function of the end-to-end distance $x = m\lambda l_0 / l_{arc}$. The front factor $m$ (Fig. 13 B) was determined by requiring the predictions of Eq.(5) agree with the experimentally observed stiffness ratios at various compositions $(c_f, c_p)$. We find that a marginal change (0.5-5%) in $x$ is sufficient to observe the range of stiffness in the modulus that we experimentally observed. This is also supported by the evidence of a minimal apparent change in the fibrin network topology as seen from the confocal images in Fig.8. It is also consistent with the mild level of stiffening that we observe in contrast to pure fibrin networks which can in principle stiffen to a much larger extent under large external deformation (by reorientation, straightening, bending, and buckling) [24,50,58].

The computed values of $m$, the efficiencies of transferred volumetric strain to the fibrin network, are plotted in Fig. 13 B as a function of the fibrin concentration. They vary within a narrow range, $m = 0.62 - 0.91$, for the full range of compositions probed in the experiments (inset shows the values of $m$ at various volume fraction of microgels and $c_f$, the numerical values of $m$ are provided in the SI table S2). It is significant that the efficiency factor $m$ is always rather large, that is, most of the strain of the shrinking particle is transferred to the surrounding fibrin network. More interesting is that $m$ is nearly constant for each particle concentration (i.e. negligible dependence on fibrin concentration $c_f$) and shows a monotonic decrease with particle concentration $c_p$. Taken together, along with the experimental observation of tension in the network at high temperature, these trends support the hypothesis of shrinking microgel particles causing tensile stress in the fibrin which creates overall stiffening due to the inherent strong nonlinearity of the fibrous network.

A remaining uncertainty of this model is to a priori predict *m,* rather than infer it from our measurements. This would require much more detailed knowledge about the interaction of microgel particles with the fibrin network. Nevertheless, we can still rationalize the trend of decreasing *m* with particle concentration by considering that finite fiber extensibility implies an upper bound on *m* (shown in Fig. 13 B). That is, the full strain of the shrinking microgel particles cannot be transferred to the surrounding network if this strain goes beyond the finite extensibility limit at $x = 1$. Finite extensibility therefore imposes $x<1$ which creates an upper bound on $\lambda_f$, which can be translated to an upper bound on *m* as

$$m < m_{max} = \frac{1}{x_0}\left(\frac{1-\phi_i}{1-\phi_f}\right)^{1/3}. \tag{6}$$

We find $m_{max}$ varies from 0.65-0.96, generally decreasing as a function of microgel particle loading as shown in Fig. 13 B (open symbols). Our experimentally inferred values of *m* are fairly close to this upper bound limit. It is possible for $m_{max} = 1$, but we never see this in our composition space due to the high volume fraction of microgel particles. In this limit, even a small volume change of particles requires a large *relative* volume change of the fibrin, and it is the relative volume change that creates the stretch $\lambda$ in Eq.(3). Designing composites such that $m_{\max} < 1$ is a strategy for dramatic stiffening, since the full theoretical de-swelling of microgel particles is beyond the finite extensibility limit. In practice, $m = m_{max}$ cannot be achieved exactly, but the fixed junction model helps elucidate how close the limit is approached in our experiments.

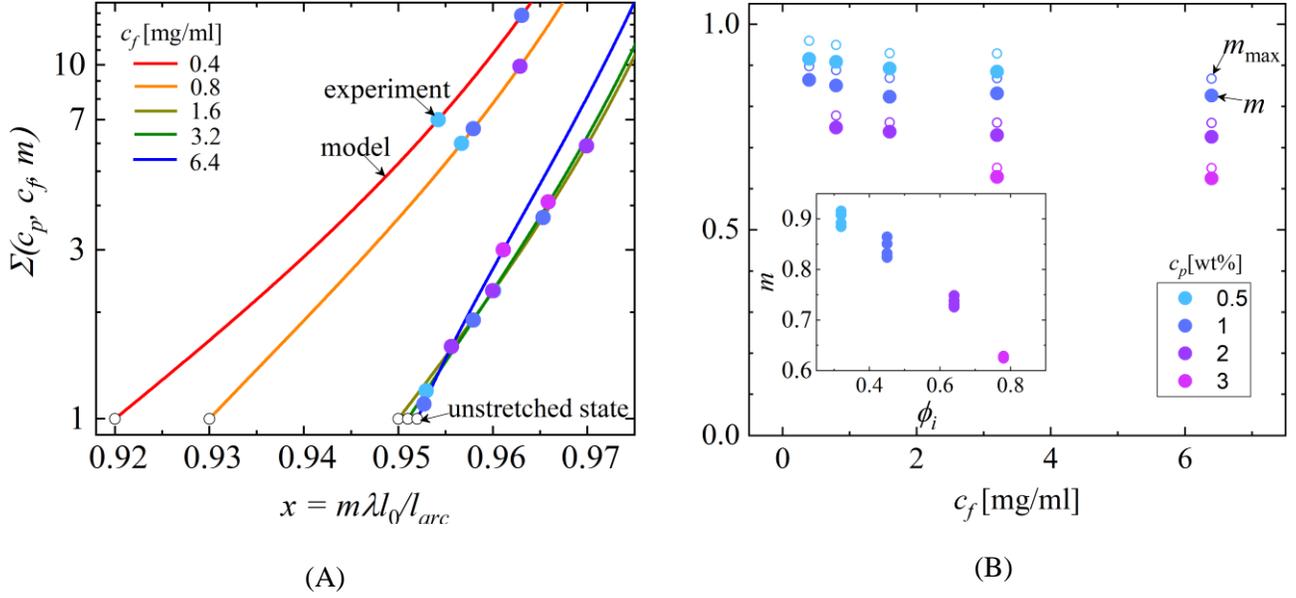

**Figure 13** (A) Model predictions (curves) of the linear shear modulus normalized by its values at $x = x_0$, . The finite extensibility of the fibers between fixed junctions leads to a drastic increase in the stiffness of the network as apparent from the plots. The symbols are experimentally measured values of $\Sigma$, and by forcing $\Sigma$ to match the model we infer $m$ for various compositions. (B) The values of $m$ (closed symbols) for which the experimental stiffening ratio and model predictions match. The maximum value of m given by Eq.(6) are plotted as open circles. Inset shows the monotonically decreasing trend of $m$ as a function of effective volume fraction of microgels.

**IV.B. Mesh adjustment model (mobile junctions)**

As an alternative distinct physical hypothesis for how the fibrin network might change when it experiences attractive interactions with de-swelling microgels we model the crosslinks as mobile physical junctions or entanglements. Indeed, the scaling of the fibrin shear modulus with concentration observed experimentally for our particle-free samples is consistent with existing models [21] of entangled semiflexible polymer liquids, rendering plausible this alternative hypothesis. A microgel-induced change of the fibrin network mesh size (entanglement length) could also be a mechanism for composite stiffening which does not necessarily involve significant deformation of fibers. In this minimalist model, we assume that the fibers can slide relative to each other and adjust their network mesh size $\xi$ and entanglement density locally without (to leading order) bending, compressing, or stretching of filaments which is energetically expensive. A schematic of this idea, which we call the "mesh adjustment model", is given in Fig. 14A.

To quantify this idea we adopt the well-known analysis for the linear shear elasticity of semiflexible entangled polymer liquids [21,24]. In the limit of infinitesimal strain, the linear elastic modulus can be written as $G \sim \frac{\rho k_B T l_p^2}{l_e^3}$. Here, the leading factor of $\rho \sim \xi^{-2}$ is a (stoichiometric) contribution to the modulus. In addition, the entanglement length is given by $l_e \sim \xi^{4/5} l_p^{1/5}$ [21,24]. Hence the linear elastic modulus can be expressed as $G \sim \rho k_b T l_p^{7/5} \xi^{-12/5}$. Note that there are two different types of filament density dependences: (a) dependence of number density of polymers that does not correlate with structural changes, and (b) one that directly impacts the force-extension behavior of individual fibers via the entanglement length $l_e$. For the simple "adjusted mesh" model described below we assume contribution (b) is the source of reinforcement.

We consider the pure fibrin polymer matrix to have an average mesh size $\xi$ and hence total network volume of $V_t = N\xi^3$, where $N$ is the total number of 3-d meshes or pores. Since the composites were synthesized at low temperature by polymerizing the fibers around the microgel, we hypothesize that the initial volume fraction of microgels determines the covered fraction of fibrin meshes, where "covered fraction" means meshes that encapsulate a single microgel. Our central hypothesis for the reinforcement mechanism then is that at the highest temperature studied in experiments beyond LCST there are strong fibrin-microgel adhesive contacts which induce the covered meshes to attain a size equal to that of a completely de-swollen microgel $D_0$. Note that this hypothesis assumes the microgels collapse in the composite to the same degree as they do in dilute solution. Such a strong collapse seems physically plausible, or at least consistent with, the idea that the fibers do not deform per the core assumption of the "adjusted mesh" model. Conservation of total volume then implies that the meshes not encapsulating a microgel must expand to compensate for the shrunken meshes. This allows us to write

$$N\xi^3 = N\phi D_0^3 + N(1-\phi)\xi_1^3 \qquad (7)$$

where $\xi_1$ is the average size of the expanded or "renormalized" mesh which is given in terms of the known average mesh size for pure fibrin ($\xi$) and the de-swollen microgel diameter $D_0$. In an effective medium spirit where a mesh is the elementary unit, we now invoke the scaling relation for shear elastic modulus described earlier to write the composite modulus as a weighted sum of contributions from shrunken meshes (which locally stiffen) and swollen ones (which locally soften):

$$\Sigma = \frac{G_{composite}}{G_{fibrin}} = \phi \left(\frac{\xi}{D_0}\right)^{12/5} + (1-\phi)\left(\frac{\xi}{\xi_1}\right)^{12/5} \qquad (8)$$

The above equation for the composite modulus above LCST is normalized to its pure fibrin value.

A caveat is that we find the above simple model gives a renormalized mesh size when the pure fibrin mesh size is smaller than the core diameter of pNIPAM microgel, or, more precisely, $\xi < \phi^{1/3} D_0$. The mean mesh size is evaluated based on the previous experimental work on similar fibrin networks (details in the SI section III). For our system, this situation applies for the highest fibrin concentration sample $c_f = 6.4$mg/ml and $c_p = 2, 3$wt% . Hence, we do not make predictions for such cases. It is difficult to know whether this is a limitation of the physical description, or simply a result of the significant experimental uncertainty in the mesh size measurements of pure fibrin, and/or the fact that we do not know precisely the mesh size *in* the composite at low temperatures where the microgel is swollen.

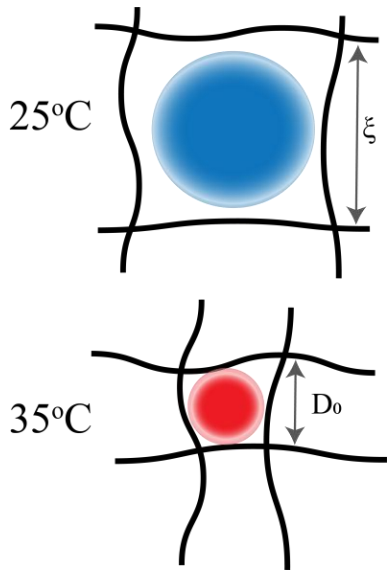
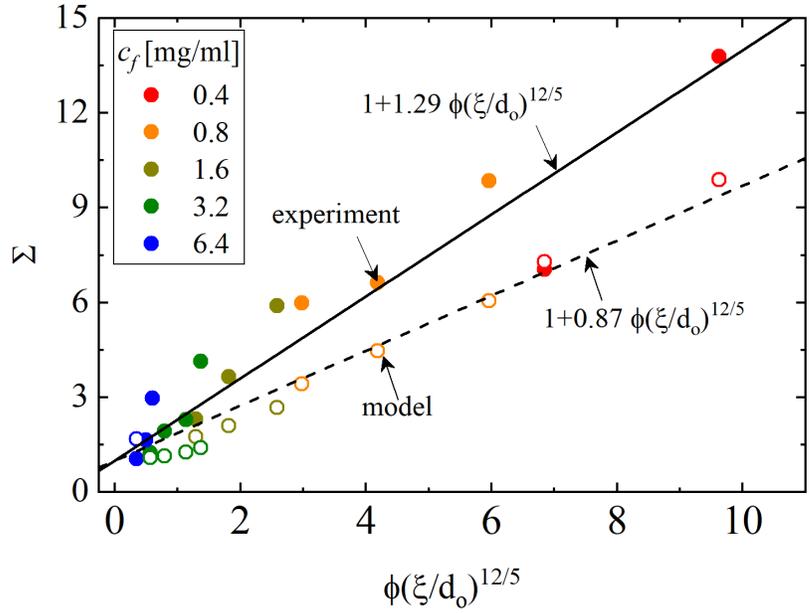

**Figure 14 Adjusted mesh model.** (A) Schematic describes the adjustment of a mesh cell that encapsulates a microgel particle as the particle de-swells beyond its LCST. (B) Comparison of experimental observations of relative stiffening with model predictions of Eq.(8)..

We now make quantitative comparisons with the experimental data to see if such a simple analysis can capture the basic trends in the experiments. Fig. 14 B compares, in a *no* adjustable or fit parameter manner, the predictions of Eq.(8) with our data. The abscissa is the dimensionless quantity that the model suggests captures the relative modulus enhancement. The value of parameters involved in this model are provided in SI table S4. We conclude that: (i) the model predictions are lower by a factor of ~1.5-2 compared with experiment (surprisingly close in our opinion), and, more importantly, (ii) the simple model does seem to capture reasonably well the qualitative trends of the experimentally observed stiffening including enhanced relative modulus with increasing microgel volume fraction and lower fibrin concentration or larger mesh size.

**V. Conclusion**

We experimentally demonstrated and theoretically modeled a new stimuli-responsive particle-polymer composite system which is able to achieve a multifold change in its shear elastic modulus under thermal excitation.

The constituents, a semiflexible biopolymer network of fibrin and thermoresponsive ultra-soft microgels, provide a novel mechanism for achieving triggerable in situ stiffening. The fundamental principle governing the responsiveness of the composite is the coupling between the strain-stiffening polymeric network of fibrin and the thermoresponsive microgels. The degree of achievable stiffness depends on the composition of both polymer and microgels. Our modeling approaches provide physical insights into the microscale mechanics of the composites and rationalizes the composition-dependent experimental trends.

Biopolymers in general perform varying functions in living organisms. Their ability to achieve a large range of mechanical properties makes them vital for many biological functions and important candidates for the development of bioinspired functional materials. Our approach highlights an effective way of harnessing the intrinsically nonlinear properties of a semiflexible polymer network into functional material design that can find applications in several potential areas such as soft robotics and biomedicine where modulating and tuning stiffness can provide novel functionality.

The nominal stiffness increase (factor of three) at the highest concentration of biopolymer and microgels also highlights the limitations in the approach that we have adopted. Biopolymer fibrin in principle can stiffen to a significantly large extent (approximately ten times the linear shear modulus) under an applied deformation, which we achieve only for the softest networks. One possible way to improve the performance of the fabricated composites and extend similar design principles to stiffer biopolymer matrices will be to either strengthen the coupling (attractions) between the microgel and the fibrous network or to fabricate microgels capable of a higher degree of volume change.

The thermoresponsive stiffening concept is not limited to biomaterial systems. Synthetic semi-flexible networks are being developed and becoming better understood [31–33]. The concept described here should in principle work with any network with strong nonlinearities where stiffness is sensitive to structural changes of the network, so long as the embedded particle free energy changes are sufficiently strong to modify the surrounding network. Our modeling approach provides a general understanding to design and engineer such composites and potentially optimize performance by a rational design of the components.

## Acknowledgement

This work was performed at the University of Illinois and supported by DOE-BES under Grant No. DE-FG02-07ER46471 administered through the Frederick Seitz 818 Materials Research Laboratory. RHE thanks Anton Paar for providing the rheometer MCR702 which was used for some of the rheology experiments.

# Supplementary Information

## I. Microgel synthesis

The lightly cross-linked monodisperse PNIPAM microgels were prepared by the surfactant-free emulsion polymerization (SFEP) method. 100 $ml$ of Type I water (18.2 $M\Omega\ cm$) was filtered through a 0.2 μ$m$ Acrodisc syringe filter. Then, 146 $mM$ (1.65 $g$) of N-isopropylacrylamide (NIPAM, 99 %, Acros) monomer was dissolved in filtered water. The monomer solution was again filtered through a 0.2 μ$m$ Acrodisc syringe filter into a 3-neck round bottom flask. The solution was stirred at 500 $rpm$, purged with nitrogen, and heated to 45°C in a temperature-controlled oil bath until the temperature of the solution became stable (1 hour typically). We then injected a solution of 2.8 $mM$ (80 $mg$) potassium peroxodisulfate (KPS, 99 %+, Sigma-Aldrich) dissolved in 1 ml of the pre-filtered Type 1 water through a 0.2 μ$m$ Acrodisc syringe filter to initiate the polymerization. The temperature was increased to 68°C roughly at a rate of 30°C/hr. The mixture was left to react under continuous stirring at 500 rpm in nitrogen atmosphere overnight. After the polymerization, the solution was cooled down to room temperature and filtered with a glass wool pad five times to remove large particulates. The microgel particles were then thoroughly purified via five cycles of a centrifuge/dispersion process. The centrifugation was done at 15000 x g of relative centrifugal force (RCF), and dispersion was enabled by a mixing process of ultrasonication followed by magnetic stirring. The cleaned particles were then lyophilized for further characterization.

## II. Average fiber length

The values of $l_0$ and $l_{arc}$ used in the effectively stretched polymer network model are obtained from literature [58] and extrapolated to estimate the values at missing concentrations (highlighted). The extrapolation was performed by assuming $l_0 = A\ c_f^{-2/5}\ l_p^{1/5}$. Table S1 gives the values of parameters used in the manuscript.

| $c_f$[mg/ml] | $l_0$ [μm] | $l_{arc}$ [μm] | $x_0$ |
|---|---|---|---|
| 0.4 | 7.17 | 8.2 | 0.92 |
| 0.8 | 5.15 | 5.7 | 0.93 |
| 1.6 | 3.8 | 4 | 0.95 |
| 3.2 | 3.1 | 3.26 | 0.951 |
| 6.4 | 2.37 | 2.49 | 0.952 |

**Table S1**. Length scales of fibrin network used in the effective stretching polymer network model

The values of $m$ for various composite compositions are given in the following table:

| $c_f$ [mg/ml] , $c_p$ [wt%] | $m$ |
|---|---|
| 0.4, 0.5 | 0.915 |
| 0.4, 1 | 0.864 |
| 0.8, 0.5 | 0.908 |
| 0.8, 1 | 0.851 |
| 0.8, 2 | 0.749 |
| 1.6, 0.5 | 0.892 |
| 1.6, 1 | 0.823 |
| 1.6, 2 | 0.738 |
| 3.2, 0.5 | 0.885 |
| 3.2, 1 | 0.832 |
| 3.2, 2 | 0.730 |
| 3.2, 3 | 0.629 |
| 6.4, 1 | 0.827 |
| 6.4, 2 | 0.726 |
| 6.4, 3 | 0.625 |

**Table S2**. Parameter $m$ extracted from the effectively stretched polymer network model

## III. Mesh size of fibrin

The mean mesh size and concentration of pure fibrin networks are related as $\xi \sim \rho^{-\frac{1}{2}}$, where $\rho$ is the fiber length density [59]. Lang et al. [59] have determined $\rho$ to be $6.1 \times 10^{11} \, m^{-2}$ for $c = 1 \, mg/ml$. We employ the known $\rho$ to estimate average mesh size of pure fibrin as,

$$\xi = \left( \frac{1}{6.1 \times 10^{11} \times c_{fibrin}} \right)^{\frac{1}{2}}$$

where $c_{fibrin}$ is in $mg/ml$. The quantitative estimates we employ are given in the table below.

| $c_f$ [mg/ml] | mesh size $\xi$ [µm] |
|---|---|
| 0.4 | 2.02 |
| 0.8 | 1.43 |
| 1.6 | 1.01 |
| 3.2 | 0.72 |
| 6.4 | 0.51 |

**Table S3.** Fibrin mesh size as a function of fibrin concentration

The various parameters involved in the adjusted mesh model are given in the following table:

| $c_f$ [mg/ml], $c_p$ [wt%] | $\xi_1$ [µm] | stiffening ratio |
|---|---|---|
| 0.4, 0.5 | 2.35 | 7.29 |
| 0.4, 1 | 2.57 | 9.9 |
| 0.8, 0.5 | 1.65 | 3.42 |
| 0.8, 1 | 1.80 | 4.46 |
| 0.8, 2 | 2.20 | 6.05 |
| 1.6, 0.5 | 1.15 | 1.75 |
| 1.6, 1 | 1.24 | 2.10 |
| 1.6, 2 | 1.5 | 2.68 |
| 3.2, 0.5 | 0.76 | 1.1 |
| 3.2, 1 | 0.81 | 1.15 |
| 3.2, 2 | 0.93 | 1.27 |
| 3.2, 3 | 1.2 | 1.40 |
| 6.4, 1 | 0.33 | 1.68 |
| 6.4, 2 | | |
| 6.4, 3 | | |

**Table S4.** Model parameters for the adjustable mesh model

## II. Elastic modulus of a single microgel particle and its deswelling in the fibrin mesh

We first model the thermoresponsive phase transition behavior of a single pNIPAM microgel in a dilute aqueous suspension. Such systems have been widely modeled using Flory-Rehner theory [37,38]. The key idea is the balance of elastic ($\pi_{elastic}$) and mixing ($\pi_{mix}$) components of osmotic pressure ($\pi$) within the particle to calculate an equilibrium degree of swelling [37]

$$\pi = \pi_{elastic} + \pi_{mix} = 0 \tag{S1}$$

$$\pi = \frac{k_B T}{\alpha^3}\left( \frac{\phi_0}{N_{gel}}\left[ \frac{\phi}{2\phi_0} - \left(\frac{\phi}{\phi_0}\right)^{1/3} \right] + \left[ -\phi - \ln(1-\phi) - \chi\phi^2 \right] \right) = 0 \tag{S2}$$

where $N_{gel}$ is the average number of monomers between crosslinks, $k_B$ is Boltzmann's constant, $\phi$ is the polymer volume fraction in a microgel, $\phi_0$ is the polymer volume fraction in the collapsed state, $\alpha$ is the monomer size, and $\chi$ is the Flory-Huggins parameter. To describe microgel collapse, a temperature and concentration dependence of $\chi$ is necessary [60]. The following form has been widely used in the literature

$$\chi = \chi_1 + \chi_2\phi + \chi_3\phi^2 + \chi_4\phi^3 \qquad (S3)$$

where $\chi_1 = 1/2 - A(1-\theta/T)$, $\theta$ is the theta temperature, $A$ depends of the mixing thermodynamics, and $\chi_2$, $\chi_3$ and $\chi_4$ are the temperature independent coefficients that account for higher order interactions between polymer and solvent molecules [60]. For the case of isotropic swelling, $\phi/\phi_0 = (d_0/d)^3$ where $d_0$ and $d$ are particle sizes in reference and current state, respectively. We use the mean hydrodynamic diameter obtained from our light scattering experiments, $d$. Hence, by using the equilibrium condition of Eq.(S2), we extract the fit parameters: $N$, $\phi_0$, $A$, $\theta$, $\chi_2$, $\chi_3$ and $\chi_4$. Fig. S1 shows the best fit curve to the DLS data. The values of fit parameter are given in table S5.

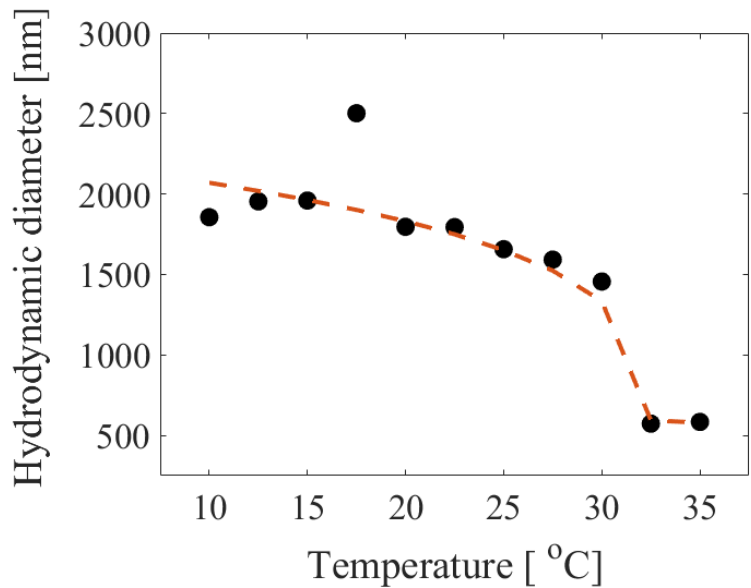

**Figure S1.** The Flory-Rehner theory fit to the DLS data appears to capture the trends in the microgel hydrodynamic size.

| | |
|---|---|
| $N_{gel}$ | 1224 |
| $\phi_0$ | 0.81 |
| $\chi_2$ | 0.30 |
| $\chi_3$ | 0.27 |
| $\chi_4$ | 0.72 |
| $A$ | -7.13 |
| $\theta$ | 305.65 |

**Table S5.** Fit parameters for the Flory-Rehner model employed.

The shear modulus for polymer networks in solvent is given by [37,40]

$$\mu = \left(\frac{\phi_0 k_B T}{2 N_{gel} \alpha^3}\right)\left(\frac{\phi}{\phi_0}\right)^{1/3} \tag{S4}$$

The parameter $\alpha$ cannot be obtained from fitting, hence we used its literature value [37] $\alpha = 6.7 \times 10^{-10} m$. This gives a shear modulus of 1.5 kPa for the fully swollen microgel particle in dilute suspension at $25^\circ C$ using the DLS data and the fit parameters described above. For evaluating the shear modulus of particle in its shrunk state above the LCST temperature we employ the classic rubber elasticity theory for cross-linked polymer networks in absence of solvent to estimate the shear elastic modulus as [37],

$$\mu = \left(\frac{k_B T}{2 N_{gel} \alpha^3}\right)\phi \tag{S5}$$

We obtain a microgel shear modulus of $\mu = 4.6 \text{kPa}$ at $35^\circ C$.

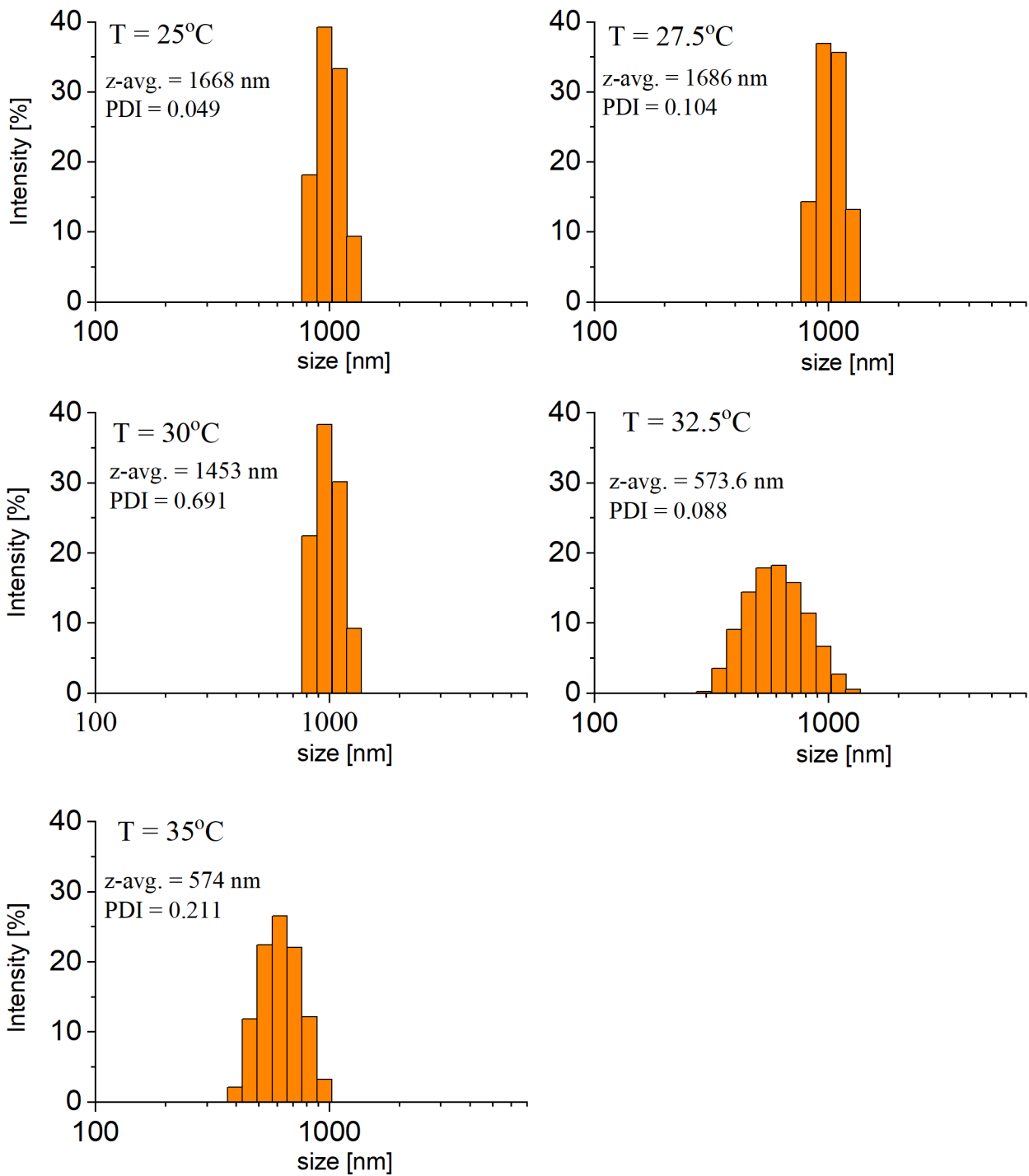

**Figure S2** Intensity versus microgel size data from the dynamic light scattering measurements (DLS) showing the distribution of particle sizes.

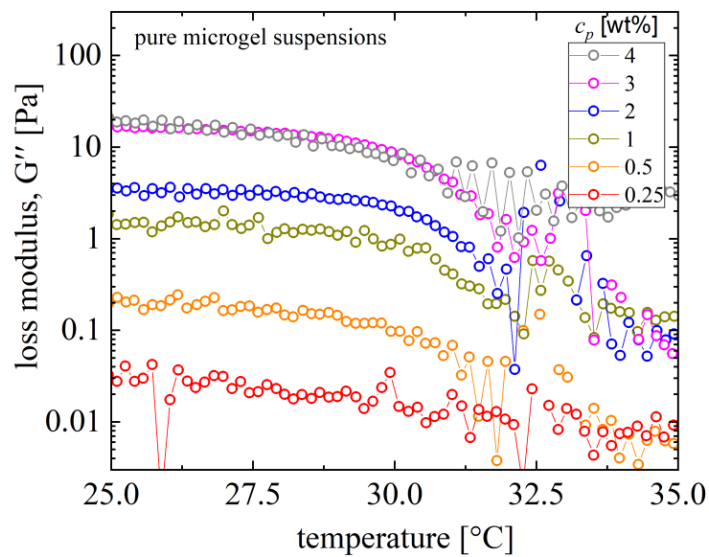

**Figure S3** The loss modulus G″ for pure pNIPAM suspensions.

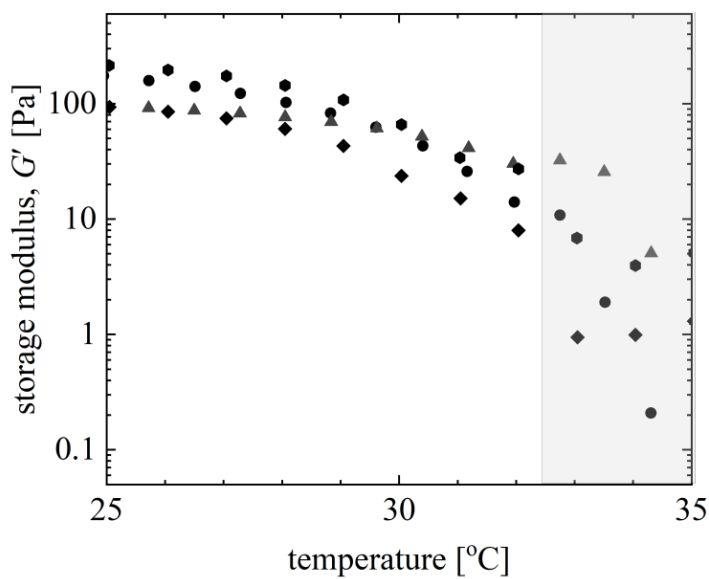

**Figure S4** Repeatability of measurements for pure pNIPAM microgel suspensions. The data sets shown are for $c_p$ = 3wt%. Above the LCST (shaded region), distinct trends can be observed in the values of *G′* because of the aggregation of microgels due to attractive interactions. Since the samples are unable to form a volume spanning network above the LCST, the inhomogeneities in the sample possibly are the origin of the variable behavior observed.

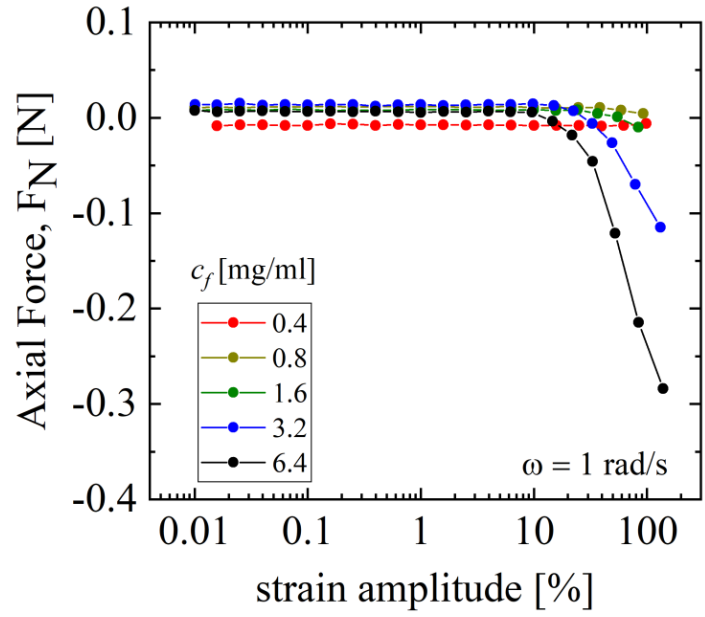

**Figure S5** Axial force for pure fibrin networks at various concentration studied as a function of applied shear strain amplitude.

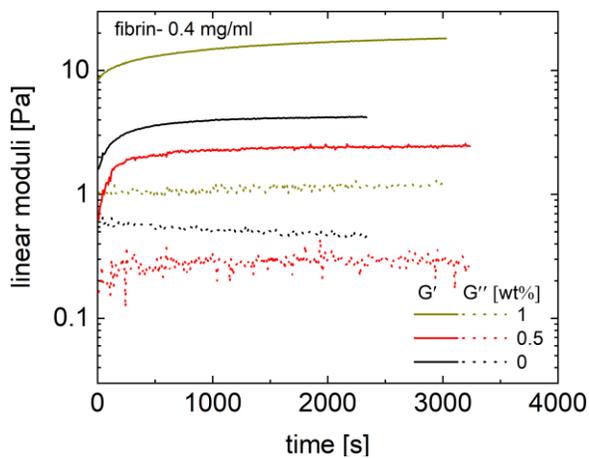

(A) 0.4 mg/ml fibrin

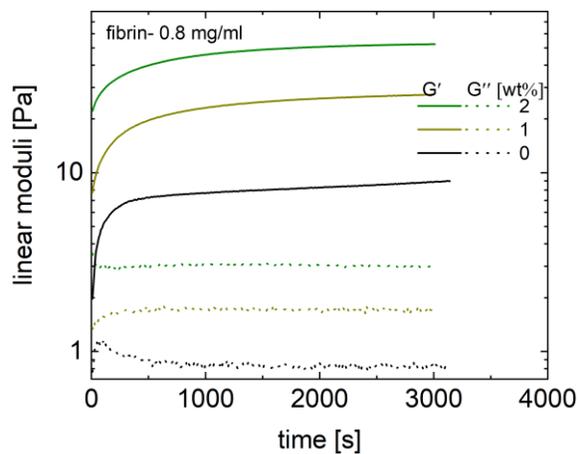

(B) 0.8 mg/ml fibrin

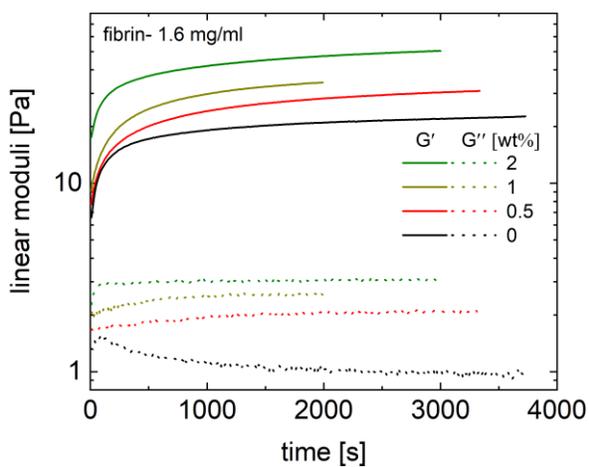

(C) 1.6 mg/ml fibrin

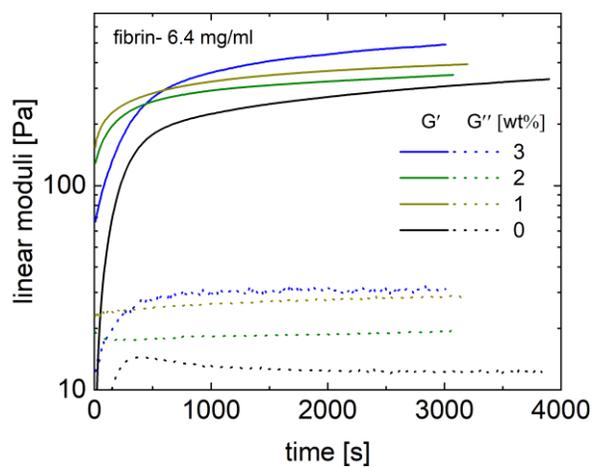

(D) 6.4 mg/ml fibrin

**Figure S6** Gelation of fibrin-pNIPAM composites. The concentration of pNIPAM microgels [wt%] used in the composites is given in the legend.

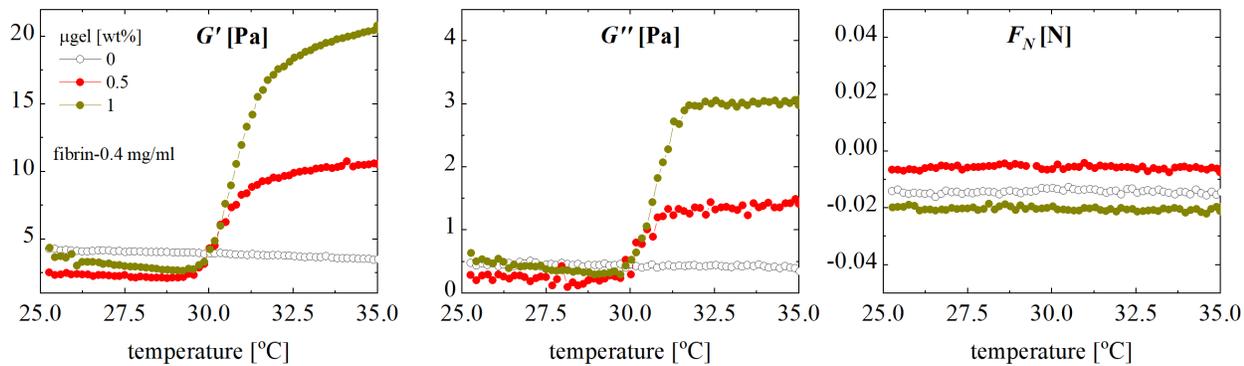

(A) 0.4 mg/ml

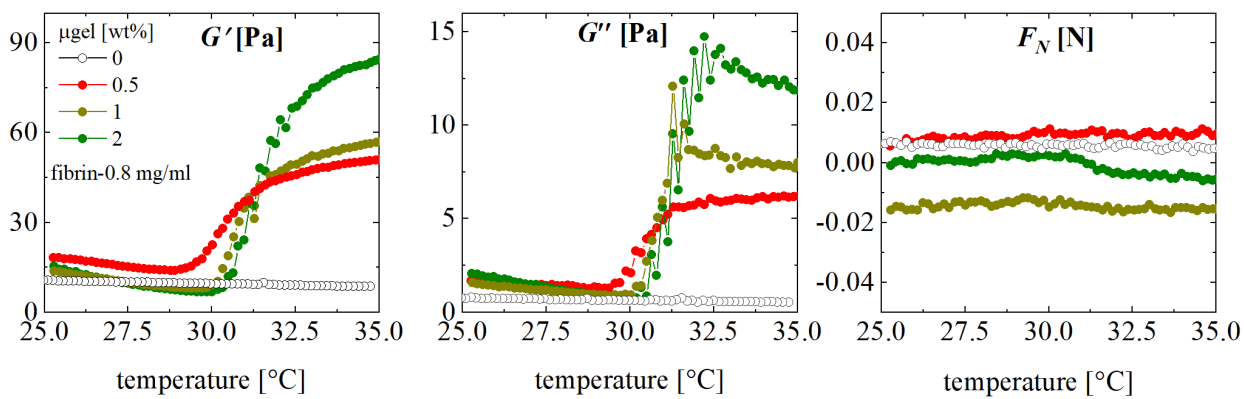

(B) 0.8 mg/ml

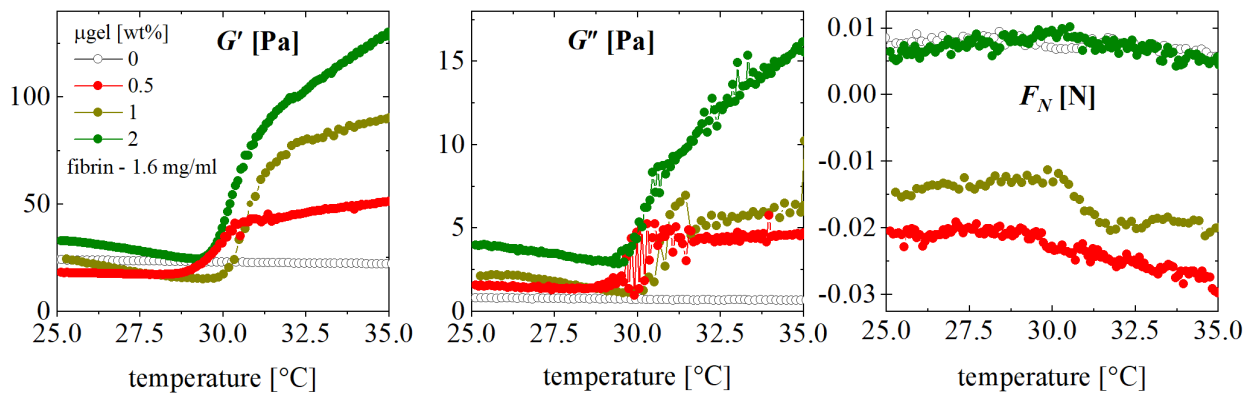

(C) 1.6 mg/ml

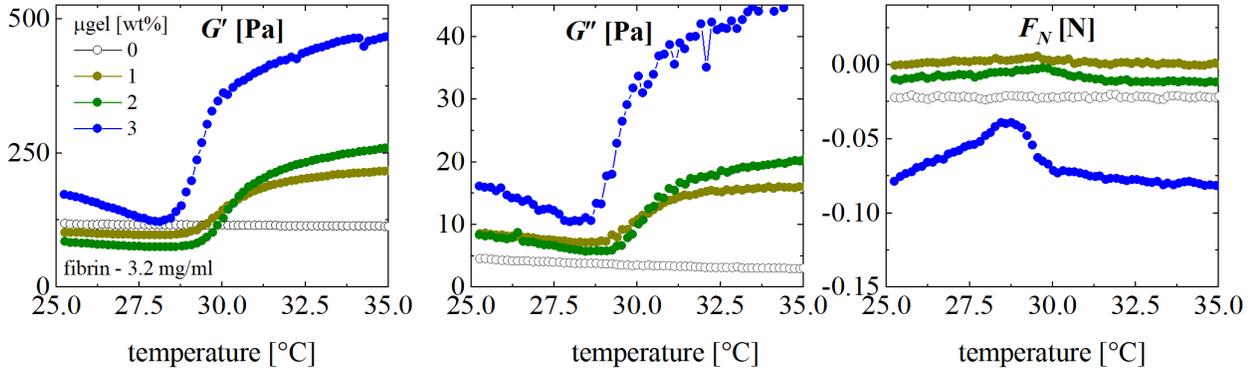

(D) 3.2 mg/ml

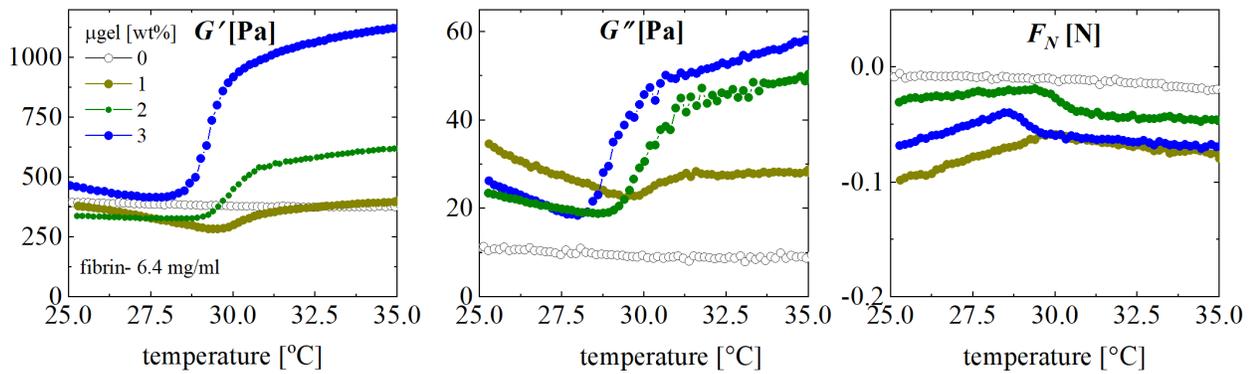

(E) 6.4 mg/ml

**Figure S7** Temperature-dependent viscoelasticity of composites with varying fibrin concentrations. The linear storage modulus $G'$, the linear loss modulus $G''$, and the axial (normal) force $F_N$ is shown for all the samples. The negative normal force indicates a downward pull on the upper plate of our rheometer.

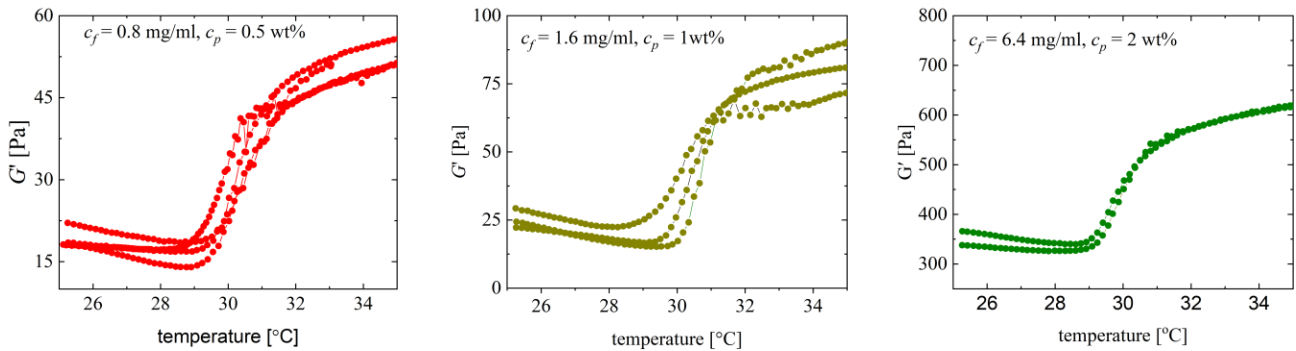

**Figure S8** Repeat measurements at selected compositions of composites.

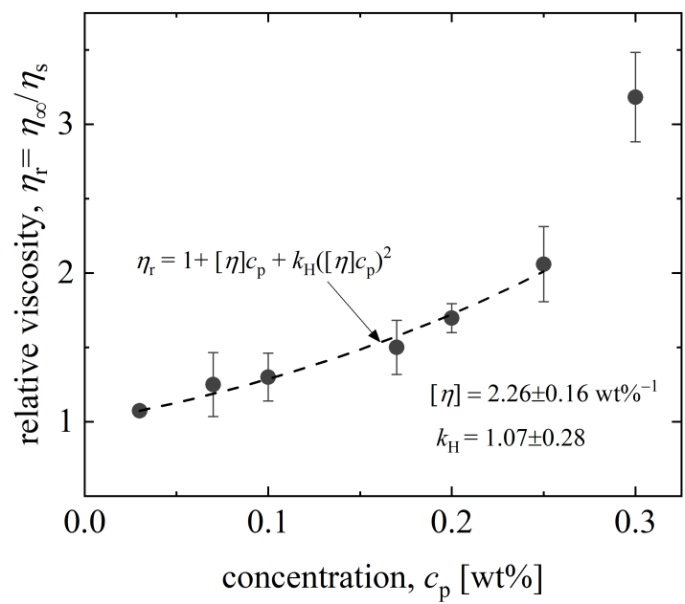

**Figure S9** Relative viscosity as a function of pNIPAM microgel wt% along with an empirical fit to the experimental data. The parameters $[\eta]$ and $k_H$ are the intrinsic viscosity and Huggins coefficient, respectively. For a dilute suspension $(c_p \to 0)$, the effective volume fraction can be related to the mass fraction as $2.5\phi = [\eta]c_p$.

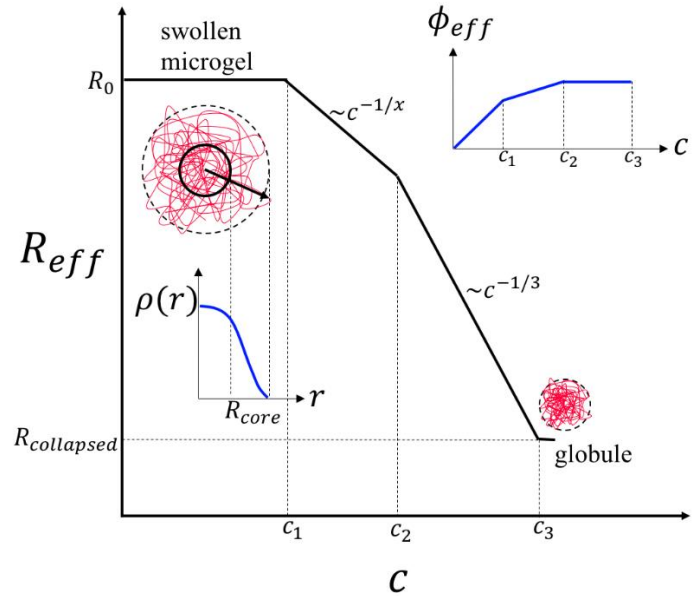

**Figure S10** Schematic of our model for microgel radius as a function of concentration (adapted from [41]), used to estimate volume fraction $\phi$ as a function of concentration $c_p$. In principle, there can be four regimes. At low concentration, the size is fixed at its $c \to 0$ dilute limit value as measured by DLS. Two intermediate regimes have different concentration dependences in the glassy fluid and "soft jammed" regimes which we envision as physically indicating first compression of the corona and then stronger shrinkage of the core due to interparticle steric repulsions. The final, perhaps not observable, regime is when the core is maximally compressed and microgel size saturates.

| $c_p$ [wt%] | $\phi_{eff,i}$ [$25^o c$] | $\phi_{eff,f}$ [$35^o C$] |
|---|---|---|
| 0.25 | 0.23 | 0.005 |
| 0.5 | 0.32 | 0.012 |
| 1 | 0.45 | 0.024 |
| 2 | 0.64 | 0.048 |
| 3 | 0.78 | 0.072 |

**Table S6.** The effective volume fraction of the microgel suspensions used in the experiments.